\documentclass[a4paper,11pt]{article}
\pdfoutput=1 % if your are submitting a pdflatex (i.e. if you have
\usepackage{jcappub} % for details on the use of the package, please
\usepackage[T1]{fontenc} % if needed

\usepackage[left]{lineno}
\title{Search for muon-neutrino emission from GeV and TeV gamma-ray flaring blazars using five years of data of the ANTARES telescope}

\author[1]{S.~Adri\'an-Mart\'inez}
\author[2]{A.~Albert}
\author[3]{M.~Andr\'e}
\author[4]{G.~Anton}
\author[1]{M.~Ardid}
\author[5]{J.-J.~Aubert}
\author[6]{B.~Baret}
\author[7]{J.~Barrios-Mart\'{\i}}
\author[8]{S.~Basa}
\author[5]{V.~Bertin}
% ...

\affiliation[1]{\scriptsize{Institut d'Investigaci\'o per a la Gesti\'o Integrada de les Zones Costaneres (IGIC) - Universitat Polit\`ecnica de Val\`encia. C/  Paranimf 1 , 46730 Gandia, Spain.}}
\affiliation[2]{\scriptsize{GRPHE - Universit\'e de Haute Alsace - Institut universitaire de technologie de Colmar, 34 rue du Grillenbreit BP 50568 - 68008 Colmar, France}}
\affiliation[3]{\scriptsize{Technical University of Catalonia, Laboratory of Applied Bioacoustics, Rambla Exposici\'o,08800 Vilanova i la Geltr\'u,Barcelona, Spain}}
\affiliation[4]{\scriptsize{Friedrich-Alexander-Universit\"at Erlangen-N\"urnberg, Erlangen Centre for Astroparticle Physics, Erwin-Rommel-Str. 1, 91058 Erlangen, Germany}}
\affiliation[5]{\scriptsize{CPPM, Aix-Marseille Universit\'e, CNRS/IN2P3, Marseille, France}}
\affiliation[6]{\scriptsize{APC, Universit\'e Paris Diderot, CNRS/IN2P3, CEA/IRFU, Observatoire de Paris, Sorbonne Paris Cit\'e, 75205 Paris, France}}
\affiliation[7]{\scriptsize{IFIC - Instituto de F\'isica Corpuscular, Edificios Investigaci\'on de Paterna, CSIC - Universitat de Val\`encia, Apdo. de Correos 22085, 46071 Valencia, Spain}}
\affiliation[8]{\scriptsize{LAM - Laboratoire d'Astrophysique de Marseille, P\^ole de l'\'Etoile Site de Ch\^ateau-Gombert, rue Fr\'ed\'eric Joliot-Curie 38,  13388 Marseille Cedex 13, France}}
% ...

\emailAdd{dornic@cppm.in2p3.fr}
\emailAdd{agustin.sanchez@ific.uv.es}

\abstract
{The ANTARES telescope is well-suited for detecting astrophysical transient neutrino
sources as it can observe a full hemisphere of the sky at all times with a high duty cycle. 
The background due to atmospheric particles can be drastically reduced, and the point-source sensitivity improved, by 
selecting a narrow time window around possible neutrino production periods. 
Blazars, being radio-loud active galactic nuclei with their jets pointing almost directly towards the observer, 
are particularly attractive potential neutrino point sources, since they 
are among the most likely sources of the very high-energy cosmic rays. Neutrinos and gamma rays may be produced 
in hadronic interactions with the surrounding medium. Moreover, 
blazars generally show high time variability in their light curves at different wavelengths and on various time scales.
This paper presents a time-dependent analysis applied to a selection of flaring gamma-ray blazars observed by the FERMI/LAT experiment 
and by TeV Cherenkov telescopes using five years of ANTARES data taken from 2008 to 2012. The results are compatible with fluctuations of the 
background. Upper limits on the neutrino fluence have been produced and compared to the measured 
gamma-ray spectral energy distribution.
}

\begin{document}
\maketitle
\flushbottom

\section{Introduction}
Neutrinos are unique messengers for studying the high-energy Universe as they are neutral, 
stable, interact weakly, and travel directly from their sources without absorption or deflection. 
Therefore, the reconstruction of the arrival directions of cosmic neutrinos would allow both the 
sources of the cosmic rays - supernova remnant shocks, active galactic nuclei jets, 
gamma-ray bursts, etc.~\cite{bib:reviewNeutrinosources} - and the relevant acceleration mechanisms 
acting within them to be identified.

The high-energy extragalactic sky is dominated by active galactic nuclei (AGN). Their spectral energy distribution can be 
described by two components: a low-energy one from radio to X-rays and a high-energy one from X-rays to very 
high-energy gamma rays. The low-energy component is generally attributed to synchrotron radiation in the relativistic 
jet by a non-thermal population of accelerated electrons and positrons; the origin of the second component is still under discussion. 
In leptonic models~\cite{bib:AGNleptonic}, it is ascribed to an inverse Compton process between the electrons and a low-energy photon field 
(their own synchrotron radiation, or external photons), while in hadronic models it originates from synchrotron emission 
by protons and secondary particles coming from p-$\gamma$ or p-p interactions~\cite{bib:hadronic1,bib:hadronic2}. Associated with these very high-energy gamma rays 
from $\pi^{0}$ decays, the decay of the charged pions gives rise to a correlated neutrino emission.

Flat-spectrum radio quasars (FSRQs) and BL Lacs, together classified as blazars, exhibit relativistic jets pointing almost directly towards
the Earth, and are among the most violent variable high-energy phenomena in the Universe~\cite{bib:Blazars}. 
Blazars are particularly attractive potential neutrino
sources, since they are among the most likely sources of the very high-energy
cosmic rays~\cite{bib:hadronic3,bib:hadronic4}. Several theoretical models predict high-energy neutrino 
emission from blazars that yield different shapes and 
normalisations for the expected energy spectrum~\cite{bib:AGNhadronic1,bib:AGNhadronic2,bib:AGNhadronic3,bib:AGNhadronic4}. In addition, the 
models also suffer from large uncertainties that originate, for instance, from 
unknowns in the model parameters, the luminosity functions and the source evolution. For example, the FSRQs are predicted to be more 
promising neutrino candidates than BL Lacs in Ref.~\cite{bib:model2}, while in Ref.~\cite{bib:model1} the 
opposite is predicted.  Several authors even estimate more optimistic spectra with spectral indexes up to one~\cite{bib:model4,bib:model5}. The $E^{-2}$ spectrum, 
generally expected from Fermi acceleration of cosmic rays in astrophysical sources, is used as a reference spectrum. 
An energy cutoff seems to be present in most sources observed in gamma rays~\cite{bib:FermicatalogueAGN}.
For these reasons, to cover the majority of the range allowed by the models accessible to the ANTARES sensitivity (see Section 2), four 
neutrino-energy spectra are tested in this analysis: $E^{-2}$, $E^{-2}\exp(-E/10~\rm{TeV})$, $E^{-2}\exp(-E/1~\rm{TeV})$ and $E^{-1}$, where $E$ is the neutrino 
energy.

In the ANTARES telescope~\cite{bib:Antares}, events are primarily detected underwater by observing the Cherenkov light induced 
by relativistic muons in the darkness of the deep sea. Owing to their low interaction probablility, only neutrinos have the ability to cross the Earth. 
Therefore, an upgoing muon is an unambiguous signature of a neutrino interaction close to the detector. The detection of cosmic neutrinos with neutrino 
telescopes is very challenging because of the small neutrino interaction cross-section and the high background of atmospheric 
neutrinos from cosmic-ray interactions in the atmosphere. To distinguish astrophysical neutrino events from background events (muons and 
neutrinos) generated in the atmosphere, energy and direction reconstructions have been used in several searches~\cite{bib:PointSource,bib:diffuse,bib:tanami}. To improve the signal-to-noise discrimination, the 
arrival time information can be used, significantly reducing the effective background~\cite{bib:MDP,bib:AntaresMicroQ}. Blazars are known to show time variability at 
different wavelengths and on various time scales~\cite{bib:FermiLATAGNvariability,bib:AGNvariability1,bib:AGNvariability2}. 
The associated neutrino emission may exhibit similar variability, and this is used in time-dependent methods to improve 
the detection probability with respect to time-integrated approaches. 

In this paper, the results of a time-dependent search for cosmic neutrino sources using the ANTARES data taken from 2008 to 2012 is presented. 
This extends a previous ANTARES analysis~\cite{bib:flare} where only the last four months of 2008 were considered. The analysis 
is applied to a list of promising blazar candidates detected in GeV gamma rays by the LAT instrument onboard the FERMI satellite, and to a list of blazar flares 
reported by TeV gamma-ray experiments (H.E.S.S., MAGIC and VERITAS). 
After a brief description of the apparatus, the data selection and the detector performances are presented in Section 2.
The point-source search algorithm used in this time-dependent analysis is explained in Section 3. The results of the GeV and TeV flare searches 
are presented in Sections 4 and 5, respectively, and discussed in Section 6.

\section{The ANTARES neutrino telescope}

The ANTARES collaboration completed the construction of a neutrino telescope in the Mediterranean Sea with the connection 
of its twelfth detector line in May 2008~\cite{bib:Antares}. The telescope is located 40 km off the Southern coast of France
(42$^\circ$48'N, 6$^\circ$10'E), at a depth of 2475 m. It comprises a three-dimensional array of 10" photomultipliers~\cite{bib:PMT}, each housed in a glass 
sphere (Optical Modules, OMs~\cite{bib:OM}), distributed along twelve slender lines anchored at the sea bottom and kept taut by a buoy at the top. The 
lines are connected to a central junction box, which in turn is connected to shore via an electro-optical cable. Since lines are subject to the sea current and can change shape 
and orientation, a positioning system comprising hydrophones, compasses and tiltmeters is used to monitor the detector geometry~\cite{bib:Positioning}. The 
main goal of the experiment is to search for neutrinos of astrophysical origin by detecting high-energy muons ($>$100~GeV) created 
by neutrino charged-current interactions in the vicinity of the detector. 

The arrival time and intensity of the Cherenkov light on the OMs are digitised into `hits' and transmitted to shore, 
where events containing muons are separated from the optical backgrounds due to natural radioactive decays and bioluminescence, 
and stored on disk. A detailed description of the detector and of the data acquisition is given 
in Ref.~\cite{bib:Antares,bib:antaresdaq}. The data used in this analysis were taken with the full detector during the period from September 6th, 
2008 up to December 31st, 2012 (54720-56292 modified Julian day). Filters are applied in order to exclude periods in which the bioluminescence-induced optical background 
was high. The resulting effective livetime is 1044 days.

Atmospheric neutrinos are the main source of background in the search for astrophysical neutrinos. These neutrinos are produced from the 
interaction of cosmic rays in the Earth's atmosphere. To account for this background, muon neutrino events are simulated using the 
GENHEN package~\cite{bib:genhen} according to the parametrisation of the atmospheric neutrino flux from Ref.~\cite{bib:Bartol}. 
An additional source of background is due to 
mis-reconstructed atmospheric muons. Downgoing atmospheric muons are simulated with the program MUPAGE~\cite{bib:mupage1,bib:mupage2}
which provides muon bundles at the detector. The full Monte Carlo chain, which includes the simulation of Cherenkov photon production 
and of the detector response, is described in Ref.~\cite{bib:MC}. 

The applied track reconstruction algorithm is the same 
as that used in the standard ANTARES point-source search~\cite{bib:PointSource}. This algorithm derives the muon track parameters that maximise a likelihood function built 
from the difference between the expected and the measured arrival times of the hits from the Cherenkov photons emitted along the muon 
track. This maximisation takes into account the Cherenkov photons that scatter in the water and the additional photons 
that are generated by secondary particles (e.g. electromagnetic showers created along the muon trajectory). 

\begin{figure}[ht!]
\centering
\includegraphics[width=0.9\textwidth]{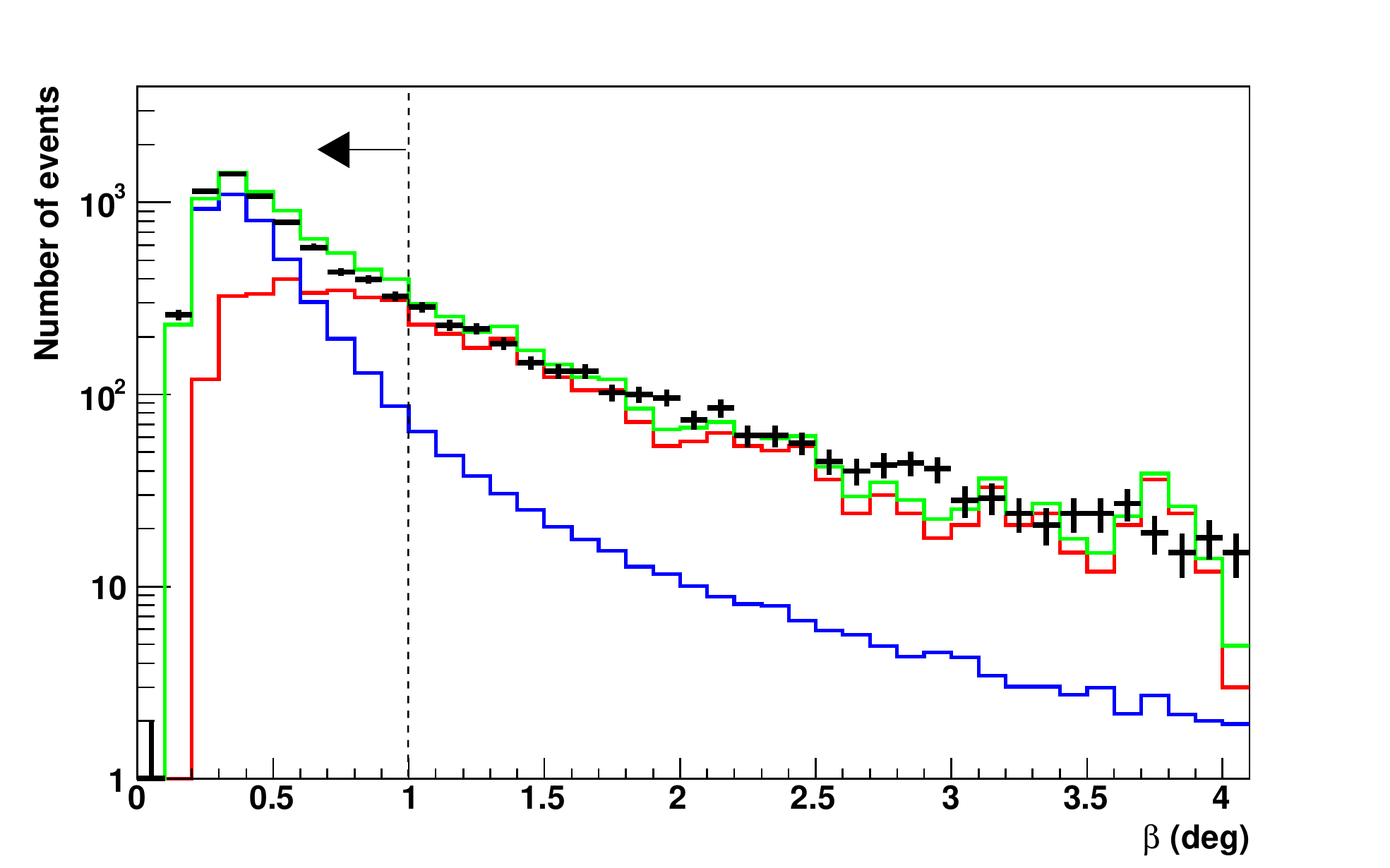}
\caption{Distribution of the estimated error on the direction of the reconstructed muon track after applying a zenith angle cut $\cos(\theta$)~$>$~-0.15 and a 
cut on the quality variable $\Lambda$~>~-5.3. The blue line shows the simulated upgoing atmospheric neutrinos, the red line the mis-reconstructed 
atmospheric muons, the green line the sum of both contributions, and the black crosses the data. The vertical dashed line with the arrow shows where the 
selection cut is applied ($\beta$~<~1$^\circ$).
}
\label{fig:beta}
\end{figure}

\begin{figure}[ht!]
\centering
\includegraphics[width=0.9\textwidth]{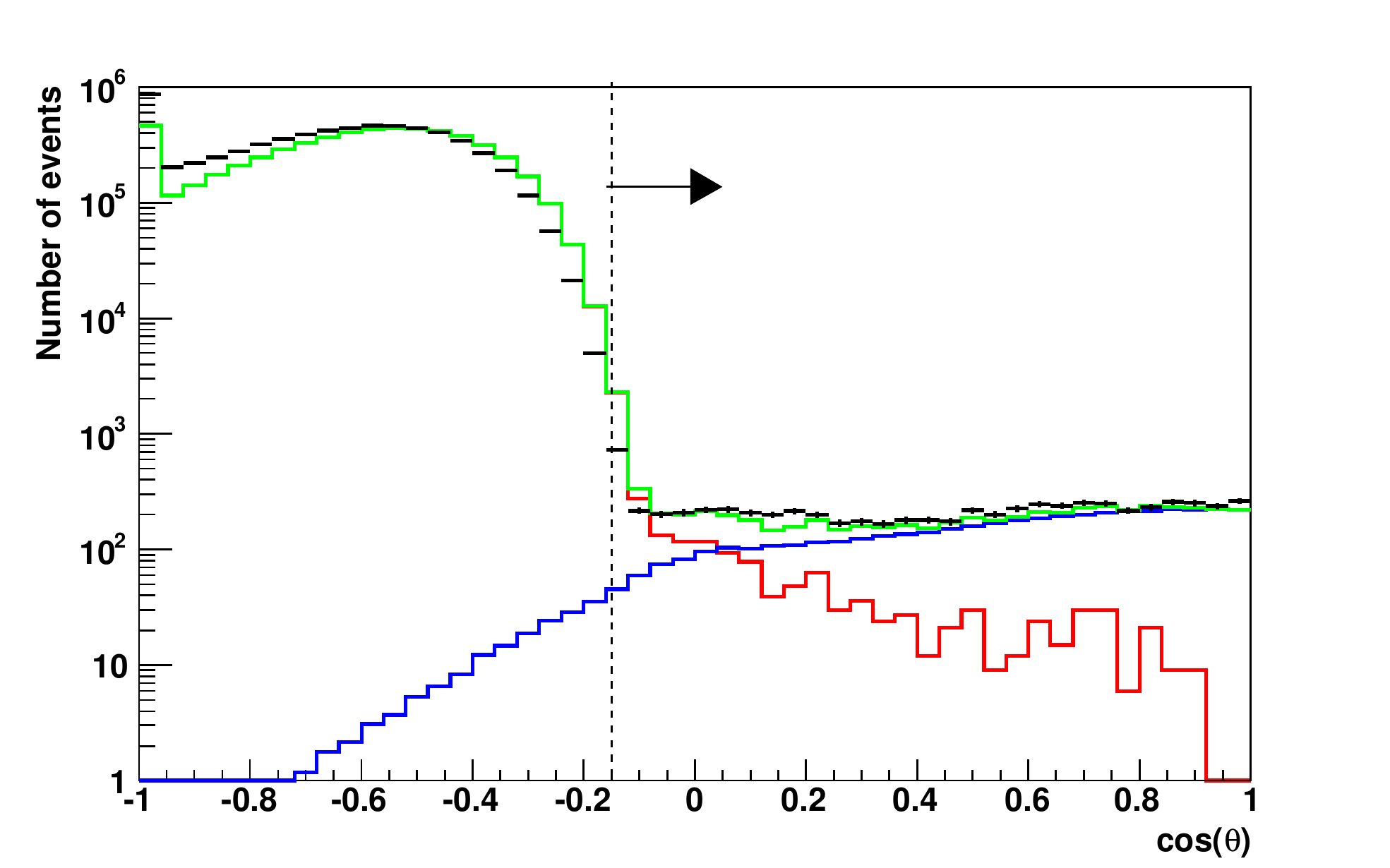}
\caption{Distribution of the reconstructed cosine of the zenith angle of the events (black crosses) with $\beta$~<~1$^\circ$ and $\Lambda$~>~-5.3. The simulated distributions 
are shown for atmospheric muons (red) and upgoing neutrinos (blue) and the green line is the sum of both components. The vertical dashed line with the arrow shows where the cut 
on the zenith angle is applied in order to select mainly upgoing events (those with larger zenith angles). }
\label{fig:costheta}
\end{figure}

\begin{figure}[ht!]
\centering
\includegraphics[width=0.9\textwidth]{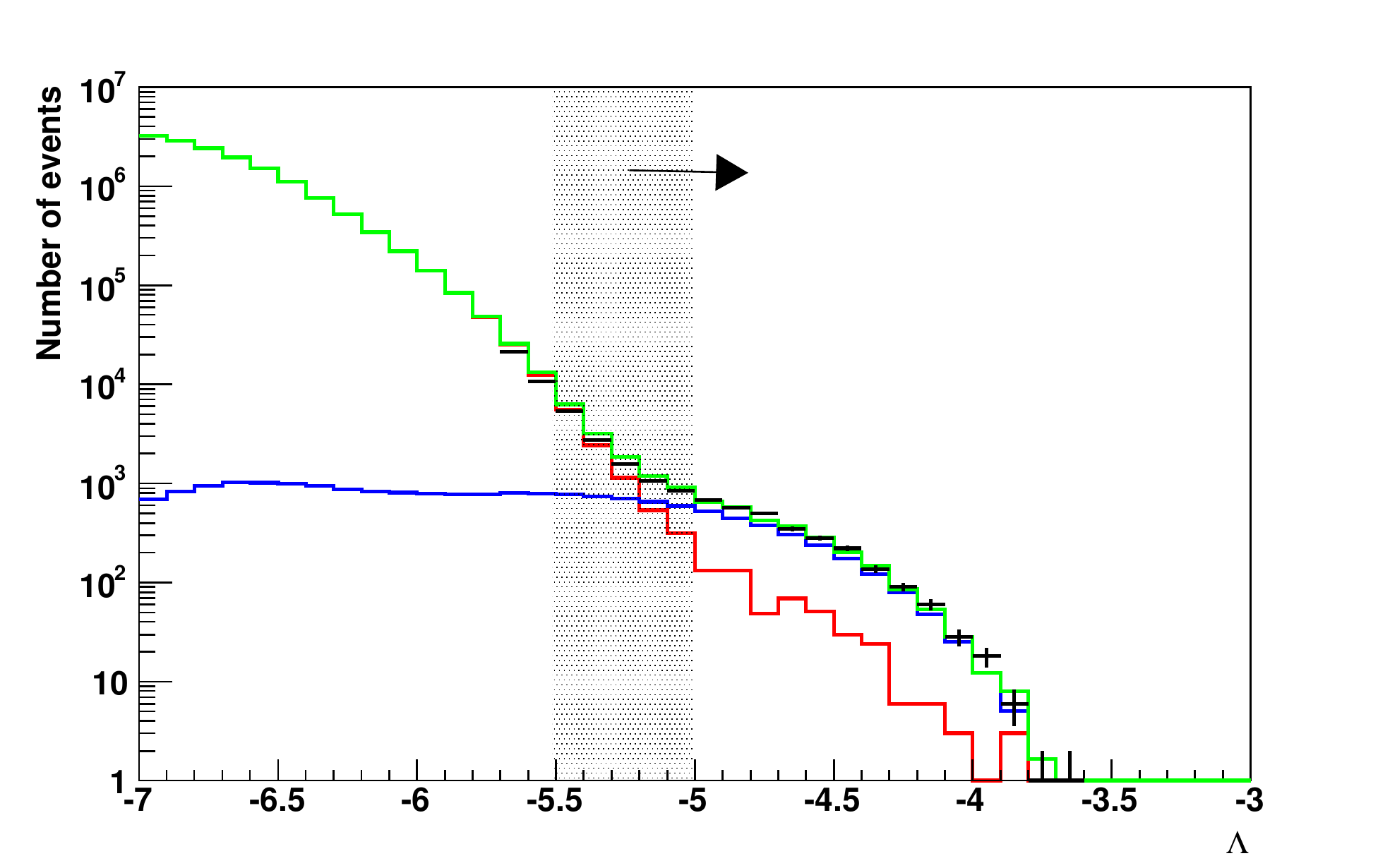}
\caption{Distribution of the reconstruction quality variable $\Lambda$ for
tracks (black crosses) which have an angular uncertainty estimate  $\beta$~<~1$^\circ$ and  zenith angle $\cos(\theta$)~$>$~-0.15. The simulated distributions 
are shown for atmospheric muons (red) and upgoing neutrinos (blue) and the green line is the sum of both components. The vertical dashed area with
the arrow shows where the optimised selection cuts stand for the various tested sources. 
}
\label{fig:lambda}
\end{figure}

The muon track reconstruction 
returns two quality parameters, namely the track-fit quality parameter, $\Lambda$, and the estimated angular uncertainty on the fitted muon track 
direction, $\beta$. Cuts on these parameters are used to improve the signal-to-noise ratio. To ensure a good directional reconstruction of the selected neutrino
candidates, $\beta$~<~1$^\circ$ is required. Figure~\ref{fig:beta} shows the distribution of the error estimate $\beta$. This cut rejects
47\% of the atmospheric muons which are mis-reconstructed as upgoing tracks.
After quality cuts, all events with a zenith angle cos($\theta$)~$>$~-0.15 are selected. Compared to a strict upgoing event selection, this cut provides a gain in visibility 
of on average 15\% for all sources with a declination above -20$^\circ$. Figure~\ref{fig:costheta} shows the distribution of the reconstructed cosine 
of the zenith angle for both data and simulation. The value of the cut on $\Lambda$ is optimised for each source on the basis of maximising a model discovery potential~\cite{bib:MDP} 
for a 3$\sigma$ significance level for each neutrino spectrum. The distribution of $\Lambda$ for events selected after applying the $\beta$ and cos($\theta$) cuts is 
shown in Figure~\ref{fig:lambda}. The optimum $\Lambda$ values 
range from -5.5 to -5.0 depending on the source and the background characteristics during the flares.

The energy of each events is estimated by exploiting the 
correlation between the energy deposition, dE/dX, and the primary energy~\cite{bib:dEdX,bib:nuspectrum}. The systematic uncertainty in the dE/dX energy estimator is 10\%.
This is accounted for by smearing the simulated signal values by a gaussian function with this RMS value.

\begin{figure}[ht!]
\centering
\includegraphics[width=0.9\textwidth]{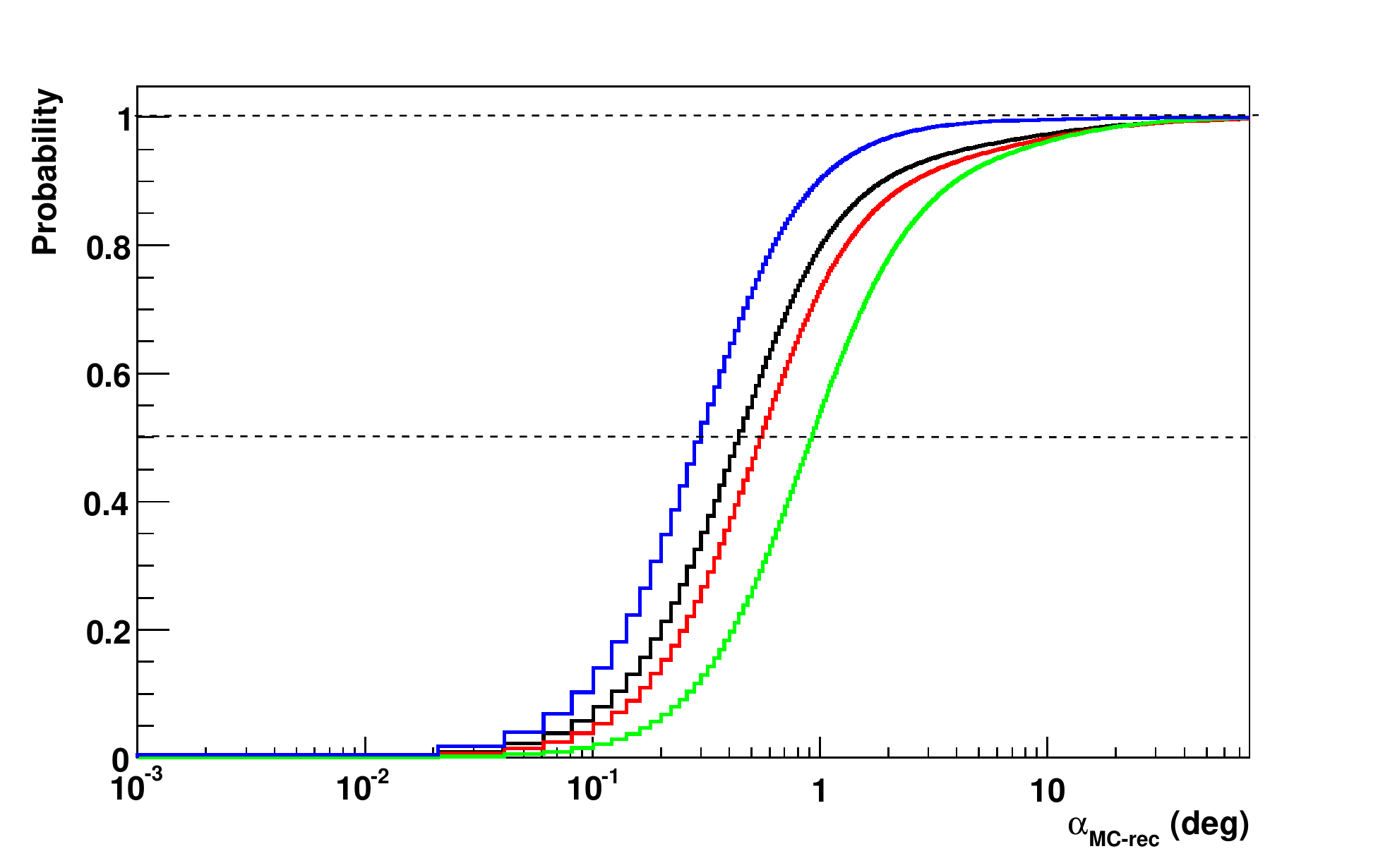}
\caption{Cumulative distribution of the angle, $\alpha$, between the true Monte Carlo neutrino direction and the reconstructed
muon direction for neutrino events selected for this analysis with $E^{-2}$ (black line, with median 0.43$^\circ$), $E^{-2}\exp(-E/10~\rm{TeV})$ (red line, with 
median 0.55$^\circ$), $E^{-2}\exp(-E/1~\rm{TeV})$ (green line, with median 0.90$^\circ$) and $E^{-1}$ (blue line,
with median 0.29$^\circ$).
}
\label{fig:Angres}
\end{figure}

The pointing accuracy has been determined directly from the data using the moon shadow~\cite{bib:Moonshadow} and is of the order of 0.63$^\circ$ (C.L. 90\%). Figure~\ref{fig:Angres} shows the cumulative distribution 
of the angular difference between the reconstructed muon direction and the neutrino direction with an assumed spectrum proportional to $E^{-2}$, 
$E^{-2}\exp(-E/10~\rm{TeV})$, $E^{-2}\exp(-E/1~\rm{TeV})$ and $E^{-1}$. For example, the median resolution is estimated 
to be 0.43$\pm$0.1$^{\circ}$ for the reference $E^{-2}$ energy spectrum. 

\begin{figure}[ht!]
\centering
\includegraphics[width=0.9\textwidth]{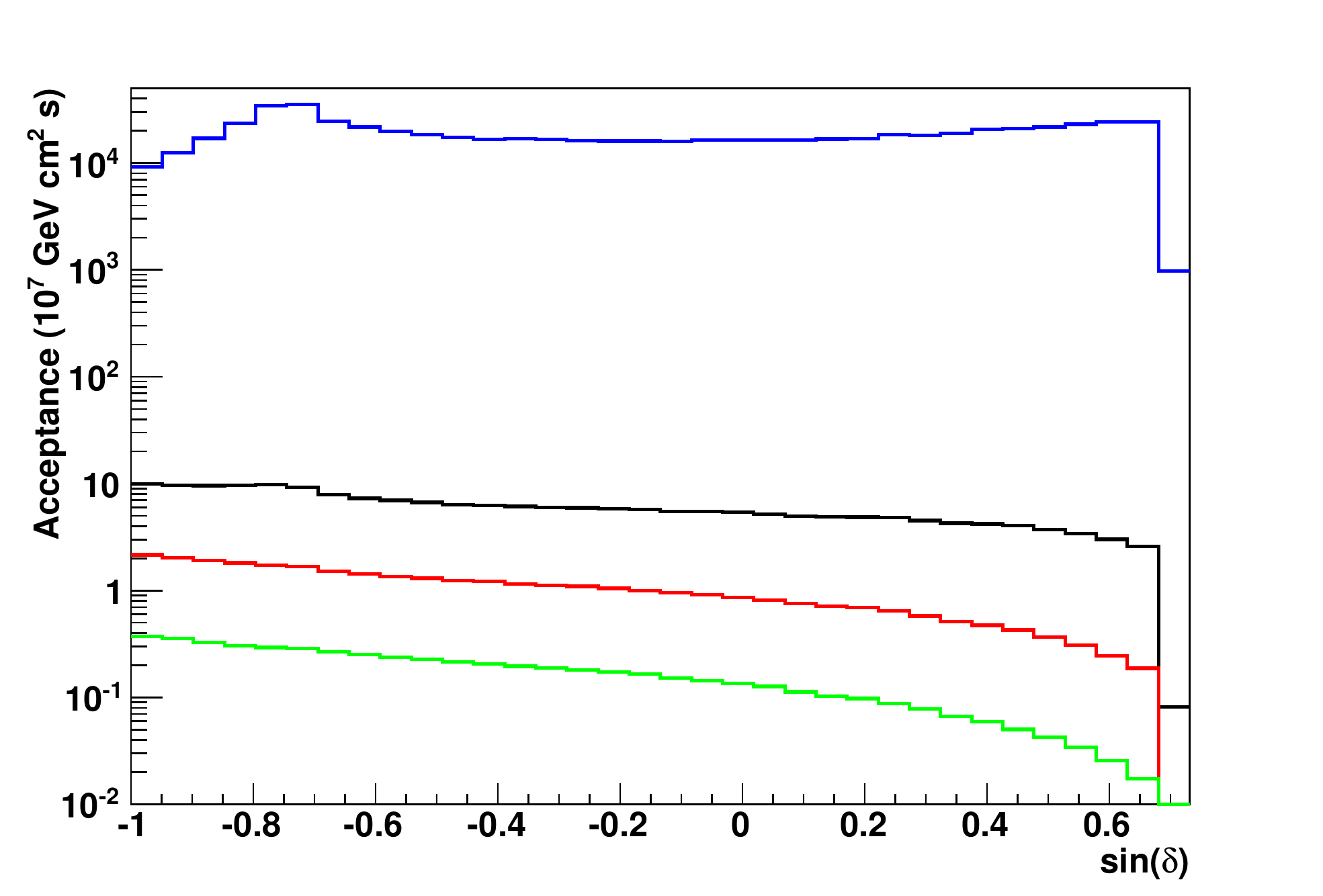}
\caption{
Acceptance of the analysis as a function of the sine of the declination for $E^{-2}$ (black), $E^{-2}\exp(-E/10~\rm{TeV})$ (red), 
$E^{-2}\exp(-E/1~\rm{TeV})$ (green) and $E^{-1}$ (blue) neutrino energy spectra. The events are selected with $\beta$~<~1$^{o}$, $\cos(\theta$)~$>$~-0.15 and $\Lambda$~>~-5.3. The 
different shape between the acceptance for $E^{-1}$ and the rest of the neutrino spectra is due to the Earth absorption for very high-energy events.
}
\label{fig:accept}
\end{figure}

The acceptance (with units of GeV~cm$^{2}$~s$^{1}$) is the proportionality factor between a given energy flux and the corresponding number of signal events expected 
in the detector after selection. Thus, a 
certain neutrino spectrum has to be assumed. For example, for $\Lambda>-5.3$, the acceptances as a function of the source declination are shown for each considered spectra in 
Figure~\ref{fig:accept}. For the limit setting, a 15\% systematic uncertainty on the acceptance is considered~\cite{bib:PointSource}.

\section{Time-dependent search method}
The time-dependent point-source analysis is performed using an unbinned method based on a likelihood-ratio maximisation. The data are 
parameterised as a two-component mixture of signal and background. The signal is expected to be small so that the full data sample (N events) can 
be treated as background. The expected number of events are $\mathcal{N}_{\rm S}$ (unknown) and $\mathcal{N}_{\rm B}$ (known) and the probabilities for 
signal and background for an event i, at time $t_i$, energy estimate $dE/dX_i$, declination $\delta_i$ and
$\Psi_i$ are $\mathcal{S}_{i}$ and $\mathcal{B}_{i}$ respectively.
$\mathcal{S}$ and $\mathcal{B}$ respectively. The probability $\mathcal{P}$ and the likelihood $\mathcal{L}$  are:

\begin{equation}
\mathcal{P}_{i} = \mathcal{N}_{\rm S}\mathcal{S}_{i} + \mathcal{N}_{\rm B}\mathcal{B}_{i}
\label{eq:EQ_likelihood1}
\end{equation}

\begin{equation}
\ln \mathcal{L} = \left(\sum_{i=1}^{N} \ln[\mathcal{N}_{\rm S}\mathcal{S}_{i}+\mathcal{N}_{\rm B}\mathcal{B}_{i}]\right)-[\mathcal{N}_{\rm S}+\mathcal{N}_{\rm B}]
\label{eq:EQ_likelihood2}
\end{equation}

\begin{figure}[ht!]
\centering
\includegraphics[width=0.9\textwidth]{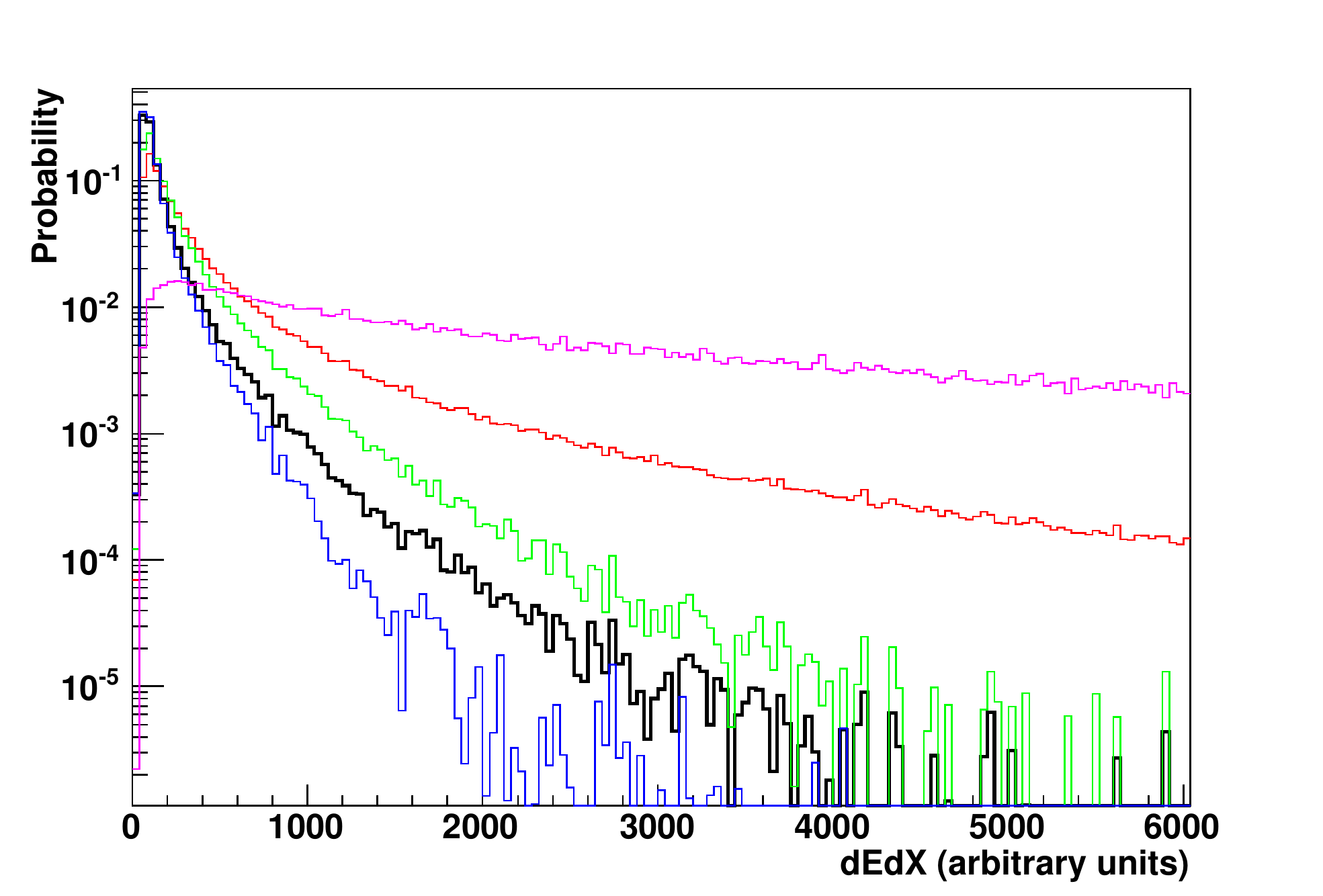}
\caption{Distribution of the energy PDF of the background events (black) and of the signal events assuming an energy spectrum following $E^{-2}$ (red), 
$E^{-2}\exp(-E/10~\rm{TeV})$ (green), $E^{-2}\exp(-E/1~\rm{TeV})$ (blue) and $E^{-1}$ (purple). Each distribution is normalised to unity.
}
\label{fig:energy}
\end{figure}

To discriminate the signal-like events from the background ones, these probabilities are described by the product of three components related to the direction, 
energy, and timing of each event. For an event \textit{i}, the signal probability is:

\begin{equation}
\mathcal{S}_{i} = \mathcal{S}^{\rm space}(\Psi_{i}(\alpha_{s},\delta_{s}))\cdot \mathcal{S}^{\rm energy}(dE/dX_{i})\cdot \mathcal{S}^{\rm time}(t_{i}+lag)
\label{eq:EQ_likelihood3}
\end{equation}

\noindent where $\mathcal{S}^{\rm space}$ is a parameterisation of the point spread function, i.e., $\mathcal{S}^{\rm space}(\Psi_{i}(\alpha_{s},\delta_{s}))$ the probability to reconstruct an 
event \textit{i} at an angular distance $\Psi_{i}$ from the true source location ($\alpha_{s}$,$\delta_{s}$). The energy PDF $\mathcal{S}^{\rm energy}$ is 
parametrised with the normalised distribution of the muon energy estimator of an event according to the studied energy spectrum. Figure~\ref{fig:energy} 
illustrates the energy PDF of these four tested energy spectra. For the generation of signal events, the correlation 
between the angular error and the energy of the muon track is taken into account. The shape of the time PDF, $\mathcal{S}^{\rm time}$, 
for the signal event is extracted directly from the gamma-ray light curve assuming the proportionality between 
the gamma-ray and the neutrino fluxes. A possible lag of up to $\pm$5 days has been introduced in the likelihood to allow for small lags in the proportionality. This corresponds
to a possible shift of the entire time PDF. The lag parameter is fitted in the likelihood maximisation together with the number of fitted signal events in the data. 

The background probability for an event \textit{i} is:

\begin{equation}
\mathcal{B}_{i} = \mathcal{B}^{\rm space}(\delta_{i})\cdot \mathcal{B}^{\rm energy}(dE/dX_{i})\cdot \mathcal{B}^{\rm time}(t_{i})
\label{eq:EQ_likelihood2}
\end{equation}

\begin{figure}[ht!]
\centering
\includegraphics[width=0.9\textwidth]{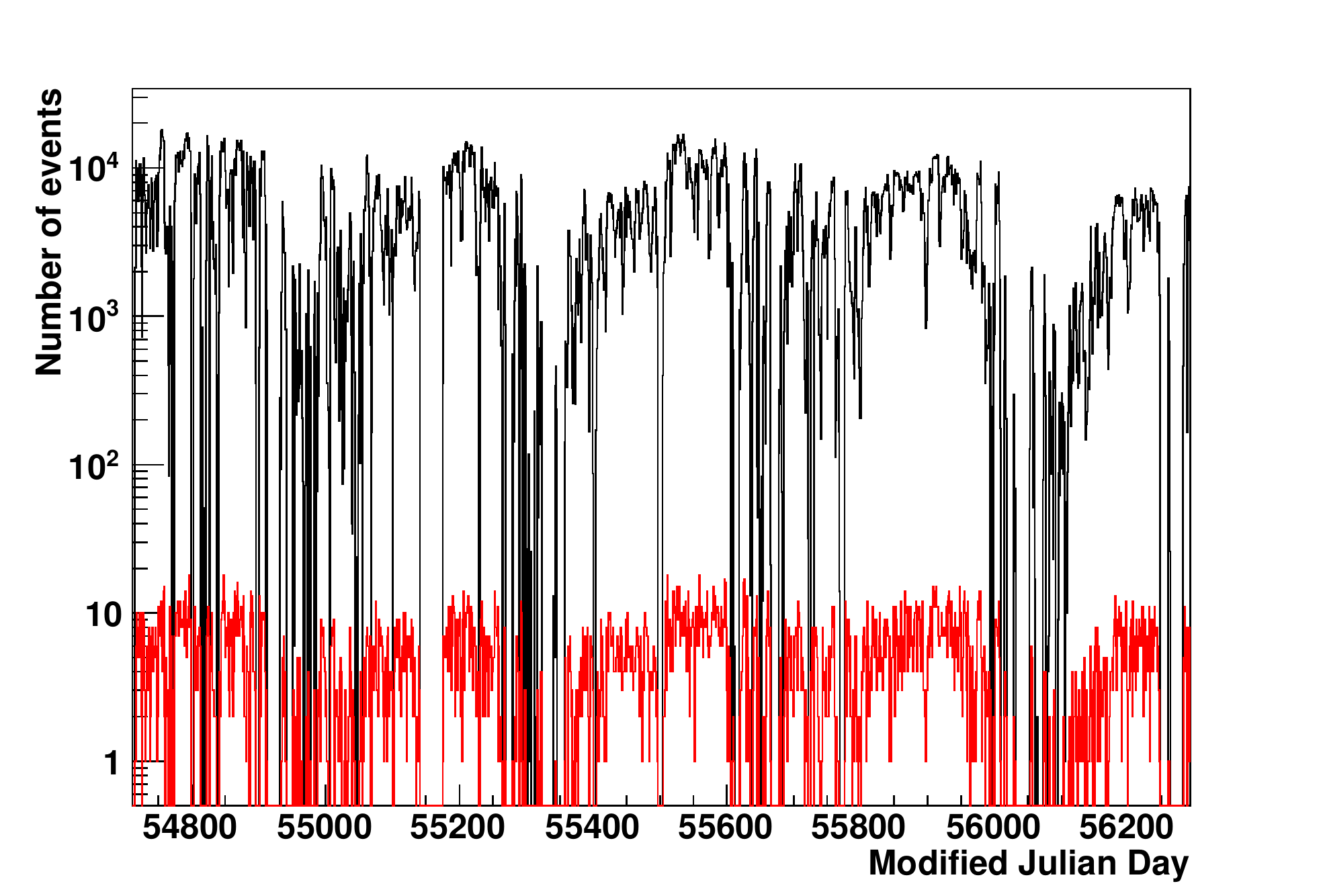}
\caption{Time distribution of the reconstructed events. Upper histogram (black): distribution of well-reconstructed events (including downgoing muons). Bottom histogram (red): 
distribution of the events selected by this analysis. 
}
\label{fig:TimeDistri}
\end{figure}

The directional PDF $B^{\rm{space}}$, the energy PDF $B^{\rm{energy}}$ and the time PDF $B^{\rm{time}}$ for the background are derived 
from data using, respectively, the observed declination distribution of selected events in the sample, the 
measured distribution of the energy estimator, and the observed time distribution of all the reconstructed muons. 
Figure~\ref{fig:TimeDistri} shows the time distribution of all reconstructed events, including both downgoing and selected events. Once normalised, the first downgoing distribution is 
used directly as the time PDF for the background. Null values indicate the absence of data taken during these periods (e.g. detector 
in maintenance) or data with a very poor quality (high bioluminescence).

The goal of the unbinned search is to determine, in a given direction in the sky and at a given
time, the relative contribution of each component, and to calculate the probability to have a signal
above a given background model. This is done via the test statistic, $\lambda$, defined as the ratio of the probability for the hypothesis 
of background and signal ($H_{\rm sig+bkg}$) over the probability of only background ($H_{\rm bkg}$):

\begin{equation}
\lambda=\sum_{i=1}^{N} \ln\frac{\mathcal{P}(x_{i}|H_{\rm{sig+bkg}}(\mathcal{N}_{\rm S}))}{\mathcal{P}(x_{i}|H_{\rm{bkg}})} 
\label{eq:TS}
\end{equation}
where $\mathcal{N}_{\rm S}$ and $N$ are respectively the unknown number of signal events and the total number of events in the considered data sample, and $x_i$ are the observed 
event properties ($\delta_i$, $RA_i$, $dE/dX_i$ and $t_i$) 

The evaluation of the test statistic is performed by generating pseudo-experiments simulating background and signal in a 30$^{\circ}$ cone 
around the considered source according to the background-only and background plus signal hypotheses. The direction of the background events are generated by 
randomly sampling the declination distribution from the background probablility $\mathcal{B}$ and the right ascension from a uniform distribution.  
Signal events are simulated by first sampling the time and the  energy of the simulated events by randomly generating from $B^{\rm time}$ and $B^{\rm energy}$, respectively. Then, the 
angular distance from the coordinates of the studied source $\Psi_{i}(\alpha_{s},\delta_{s})$ is sampled as a function of the declination and of the estimated energy.

The null hypothesis is given by $\mathcal{N}_{\rm S}=0$ ($\lambda_{\rm 0})$ (i.e. the background-only hypothesis $H_{\rm bkg}$). The obtained value of $\lambda$ for the data, $\lambda_{\rm data}$, is then 
compared to the distribution of $\lambda_{\rm 0}$ obtained by pseudo-experiments. Large values of $\lambda_{\rm data}$ compared to the 
distribution of $\lambda_{\rm 0}$ reject the null hypothesis with a confidence level depending on the fraction 
of the $\lambda_{\rm 0}$ distribution above $\lambda_{\rm data}$. This fraction of trials above $\lambda_{\rm data}$ is referred to as the p-value. 
The discovery potential is then defined as the average number of signal events required to achieve a p-value lower than $2.7~10^{-3}$ ($5.7~10^{-7}$) (3(5)$\sigma$) 
in 50$\%$ of the trials.

\begin{figure}[ht!]
\centering
\includegraphics[width=0.9\textwidth]{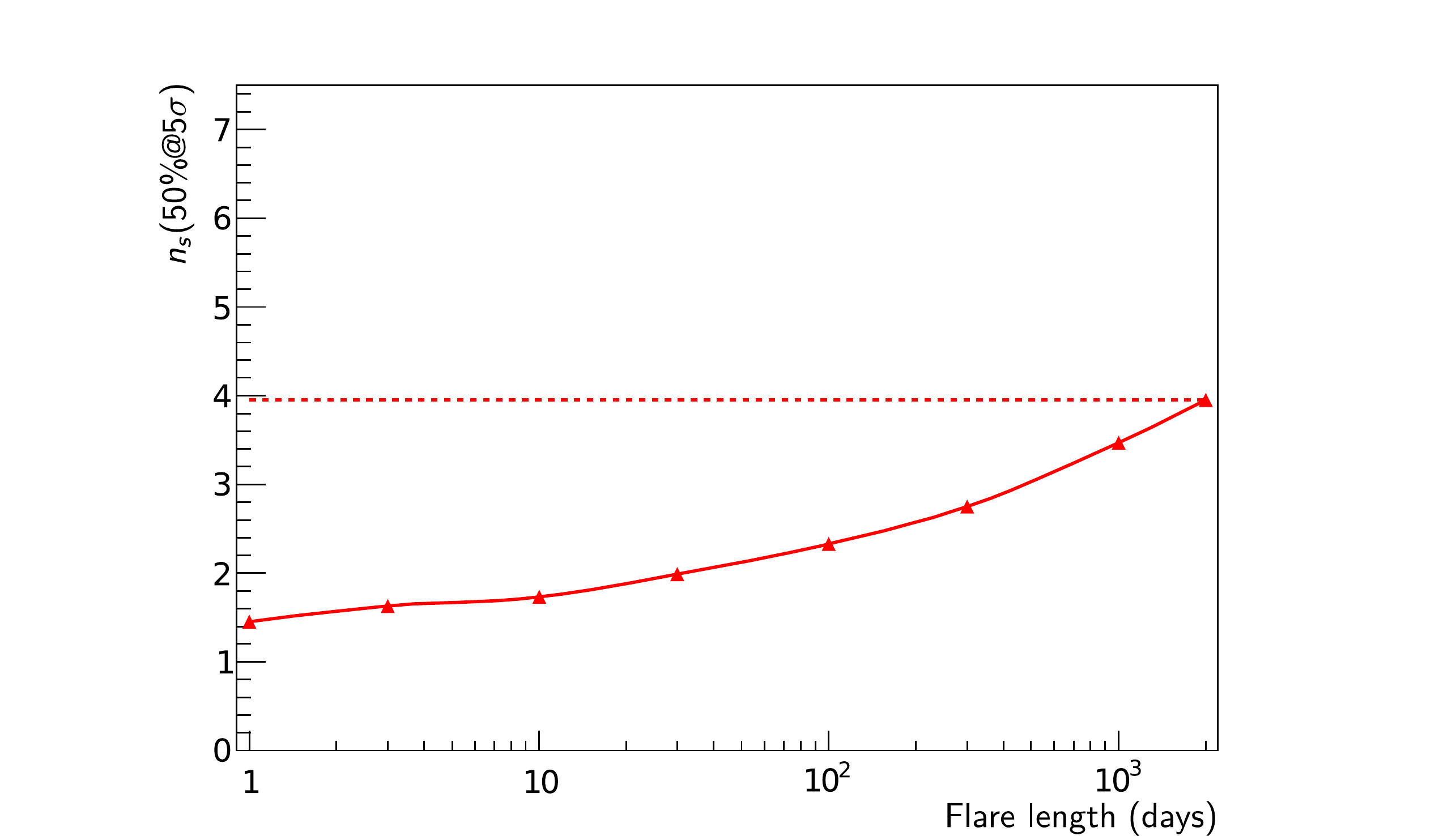}
\caption{Average number of events required for a 5$\sigma$ discovery (50$\%$ probability) for a source located at a
declination of -40$^{o}$ and following an $E^{-2}$ energy spectrum as a function of the total width of the flaring periods (solid line). These
numbers are compared to those obtained without using the timing information (dashed line).}
\label{fig:Nev3sigma}
\end{figure}

The performance of the time-dependent analysis is computed with a toy experiment with a source assuming a 
square-shaped flare with a width varying from 1 to 2000 days assuming a flat background period of 2000 days. Figure~\ref{fig:Nev3sigma} 
shows the average number of events 
required for a 5$\sigma$ discovery for a single source located at a declination of -40$^{o}$ and following an $E^{-2}$ energy spectrum, as a function of the total width 
of the flare periods. These numbers are compared to those obtained without the selection of time intervals corresponding to flaring periods. For time ranges 
characteristic of flaring activity, the time-dependent search 
presented here improves the discovery potential by on-average a factor 2-3 with respect to a standard time-integrated point-source 
search~\cite{bib:PointSource} under the assumption that the neutrino emission is correlated with the gamma-ray flaring activity.

\section{Search for neutrino emission from gamma-ray flares detected by FERMI}

The time-dependent analysis described in the previous section is applied to bright and variable Fermi blazar sources reported in the second Fermi LAT 
catalogue~\cite{bib:Fermicatalogue} and in the LBAS catalogue (LAT Bright AGN sample~\cite{bib:FermicatalogueAGN}). The sources 
located in the part of the sky visible to ANTARES ($\delta~<~35^\circ$) with a flux greater than $10^{-9}~\rm{photons}\cdot \rm{cm}^{-2}\cdot \rm{s}^{-1}$ 
above 1~GeV, a test statistic $TS~>~25$ (corresponding to a detection significance of more than 4 sigma) and a significant 
time variability are selected. This list is completed by adding sources reported as flaring in the Fermi Flare Advocates in 2011 
and 2012~\cite{bib:FermiAdvocates}. The final list includes a total of 153 sources. 

\begin{figure}[ht!]
\centering
\includegraphics[width=1.0\textwidth]{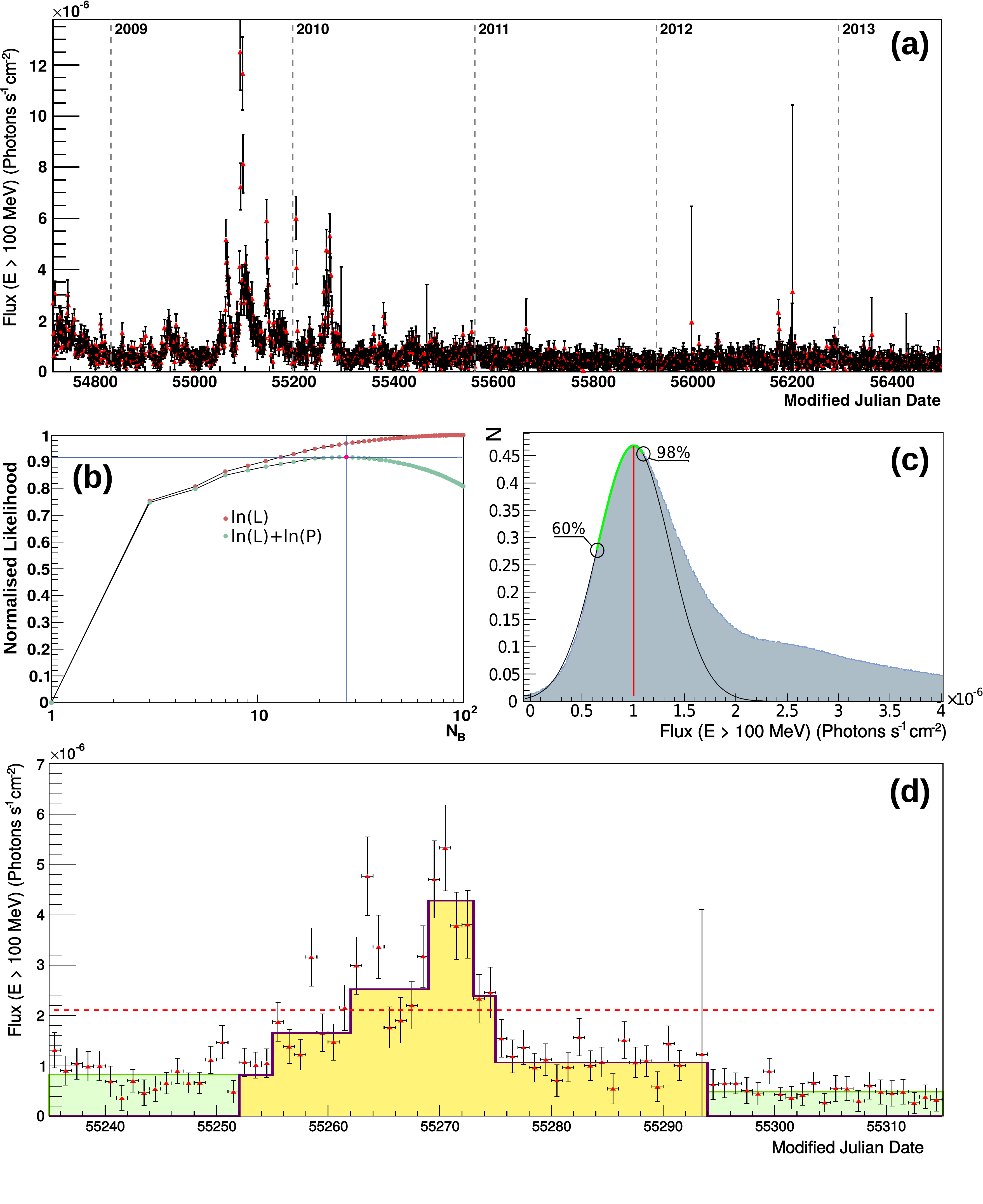}
\caption{Illustration of the flare-selection concept using the maximum-likelihood block algorithm with the blazar 3C273. (a) Gamma-ray light
curve using the 2008-2012 Fermi data with thershold of 100 MeV. (b) Evolution of the likelihood L (red dots) as a function of the number of
added blocks ($N_{B}$). The green dots show the evolution after including the prior $\mathcal{P}$. The solid lines indicate the selected maximum. (c) 
Histogram of the flux convolved with a Gaussian distribution of the error on the flux. A Gaussian fit, shown by the black line, has been performed 
using data over the range (illustrated by the green line) between 60\% (on the left-hand side) and 98\% (on the right-hand side) of the maximum value.
The  value of the baseline is indicated by the red line. (d) Example of a selected flare of 3C273. Dots, the green and the yellow histograms display the raw
data, the denoised light curve and the selected flare. The red dashed line corresponds to the 5$\sigma$ threshold. 
}
\label{fig:selection}
\end{figure}

Light curves for the selected sources are produced using the Fermi Public Release Pass 7 data using the source class event selection (evclass=2) and the 
Fermi Science Tools v9r35p1 package~\cite{bib:FermiData}. Following the standard event selection cuts proposed by the Fermi-LAT Collaboration~\cite{bib:FermiAna}, the data 
are filtered using the \textit{gtselect} tool to select only events 
which are most likely gamma-rays. Light curves are computed from the photon counting data in a cone of two-degree radius around each source 
direction, corrected by the total exposure. With this method the diffuse background 
contributions are not subtracted. This limits the validity of this method to only bright gamma-ray sources, but does not affect the flare identification in these
sources. Therefore, sources which are close 
to the galactic plane (galactic latitude <~10$^\circ$) or have others sources within a two-degree cone (or 3$^\circ$ for very bright sources) are excluded. 
The baseline is then removed for the time PDF definition.
The resulting light curves correspond to the one-day-binned time evolution of the average gamma-ray flux above a threshold of 100~MeV from August 
2008 to December 2012. This method has the main advantage of producing continuous and complete gamma-ray light curves. Figure~\ref{fig:selection} (a) shows 
the resulting light curve for the source 3C273. This aperture photometry method agrees with the results 
of the likelihood analysis (an alternative method used for the LAT data analysis). The exposure is calculated using the \textit{gtexposure} tool, which is also part of the Fermi framework.

A maximum likelihood block (MLB) algorithm~\cite{bib:scargle, bib:scargle1, bib:scargle2} is used to remove noise from the light curves by iterating over 
the data points and selecting 
periods during which data are consistent with a constant flux within statistical errors. 
The description of the light curve in terms of $N_{B}$ periods of constant emission is obtained with the following likelihood:

\begin{equation}
  \ln L = \displaystyle \sum_{n_{B}\in N_{B}} \frac{\left(\displaystyle \sum_{i\in n_{B}} \frac{x_{i}}{\sigma_{i}^{2}} \right)^{2}}{\displaystyle \sum_{i\in n_{B}} \frac{1}{\sigma_{i}^{2}}}
\end{equation}

where $x_{i}$ is the flux measurement of the data point \textit{i} and $\sigma_{i}$ is the error of the 
flux measurement \textit{i}. Since 
this likelihood would maximise when there are as many periods as data points, while only a few periods are required to
describe the light curve, the following prior $\mathcal{P}$ is added to the likelihood:
\begin{equation*}
  \ln \mathcal{P} = - N_{B} \ln \mathcal{N}
\end{equation*}
where $\mathcal{N}$ is the number of data points in the light curve. The algorithm searches in each iteration for the best two flux changes in the emission in each period already found 
and keeps them for the next iteration. The Figure~\ref{fig:selection} (b) shows the optimisation of the likelihood in the case of 3C273.

\begin{figure}[ht!]
\centering
\includegraphics[width=0.9\textwidth]{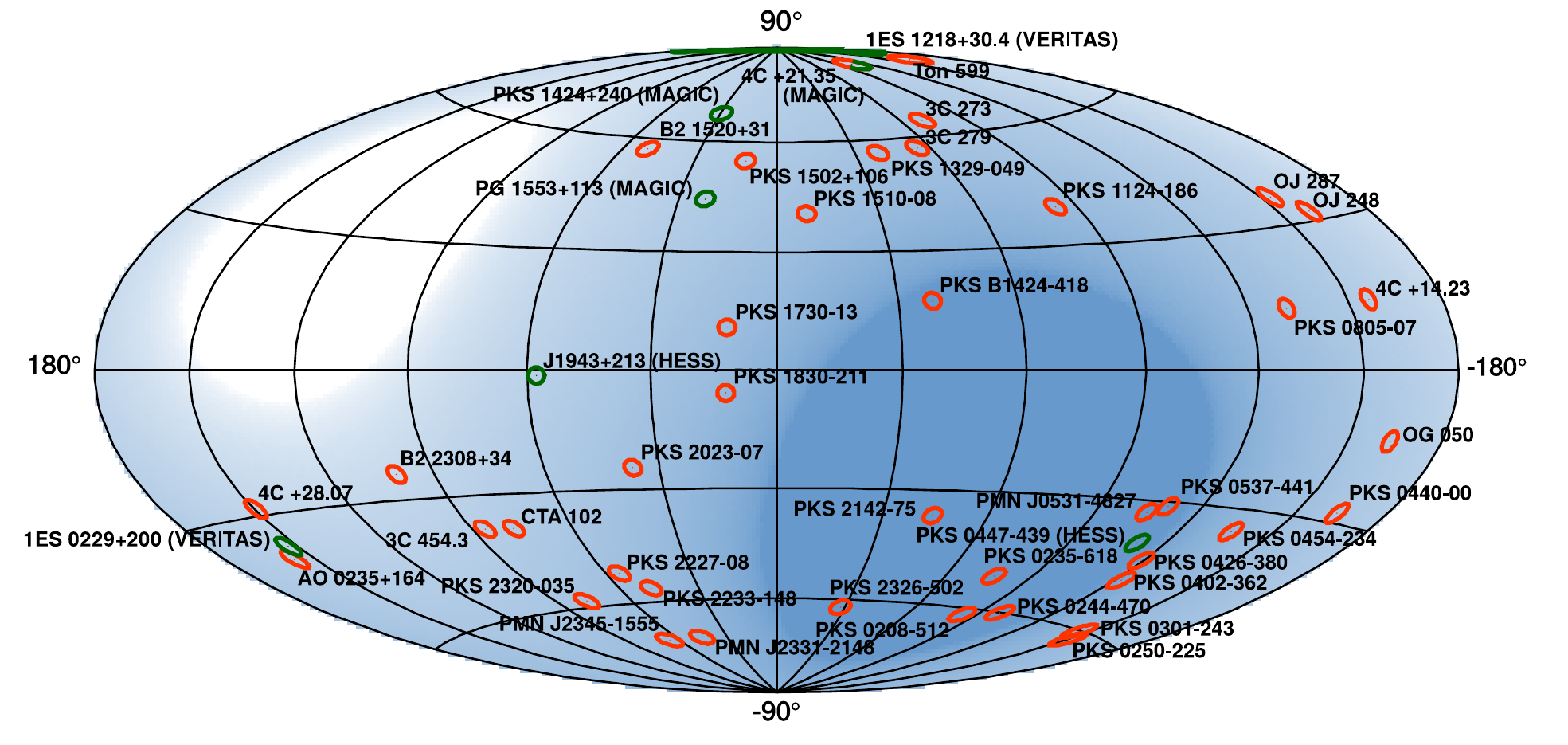}
\caption{Skymap in galactic coordinates showing the position of the 41 selected Fermi blazars (red circles, see Section 4) and the seven TeV blazars (green circles, see
Section 5) together with the ANTARES visibility (dark blue is maximal).
}
\label{fig:skymap}
\end{figure}

High-state periods are defined using a simple and robust method. The value of the steady state (i.e. baseline, BL) and its 
fluctuation ($\sigma_{BL}$) are determined with a Gaussian fit of the lower part of the distribution of the flux data points (Figure~\ref{fig:selection} (c)). 
The flaring periods are defined in three main steps. 
Firstly, seeds are identified by searching for points with an amplitude, or blocks with a fluence above $BL~+~5\sigma_{BL}$. Then, each period is extended forward and backward 
up to an emission compatible with $BL~+~1\sigma_{BL}$. An additional delay of 0.5 days is added before and after the flare in order to take into account that the precise time 
of the flare is not known (one-day binned light curve). 
Finally, spurious flares are discarded by requiring that each flare is visible in the three light curves produced 
with a gamma-ray threshold of 100, 300 and 1000~MeV. Figure~\ref{fig:selection} (d) shows an example of one flare of 3C273. 
With the above definition, a flare has a width of at least two days. Under the hypothesis that the neutrino emission follows the gamma-ray emission, the signal time PDF is the normalised 
de-noised light curve with only the high state periods (the other periods are set to zero). The final list includes 41 bright and variable 
Fermi blazars: 33 FSRQs, 7 BL Lacs and 1 unknown identification. Figure~\ref{fig:skymap} shows the position of the selected Fermi Blazars 
together with the ANTARES visibility. The main characteristics of these blazars are reported in Table~\ref{table:Sources}.

\begin{table}[ht!]
\caption{List of bright variable Fermi blazars with significant flares selected for this analysis.}	       
\label{table:Sources}	 
\centering
\begin{tabular}{|c ||c |c |c |c |c|}
\hline Name & {OFGL name} & Class & {RA [$^\circ$]} & {Dec [$^\circ$]} & Redshift \\
\hline
\hline{3C 454.3} & {J2254.0+1609} & FSRQ & 343.50 & 16.15 & 0.859 \\
\hline{4C +21.35} & {J1224.9+2122} & FSRQ & 186.23 & 21.38 &  0.434\\
\hline{PKS 1510-08} &  {J1512.7-0905} & FSRQ & 228.18 & -9.09 & 0.360 \\
\hline{3C 279} & {J1256.1-0548} & FSRQ & 194.03 & -5.8 & 0.536 \\
\hline{PKS 1502+106} & {J1504.3+1029} & FSRQ & 226.10 & 10.49 & 1.839 \\
\hline{PKS 2326-502} & {J2329.2-4956} & FSRQ & 352.32 & -49.94 & 0.518 \\
\hline{3C 273} & {J1229.1+0202} & FSRQ & 187.28 & 2.05 & 0.158 \\
\hline{AO 0235+164} & {J0238.6+1636} & BLLac & 39.65 & 16.61 & 0.940 \\
\hline{PKS 0426-380} & {J0428.6-3756} & FSRQ & 67.17 & -37.93 & 1.110 \\
\hline{4C +28.07} & {J0334.3-3728} & FSRQ & 53.58 & -37.47 & 1.206 \\
\hline{PKS 0454-234} &  {J0457.1-2325} & FSRQ & 74.28 & -23.43 & 1.003 \\
\hline{PKS 1329-049} & {J1332.0-0508} & FSRQ & 203.01 & -5.14 & 2.150 \\
\hline{PKS 0537-441} & {J0538.8-4405} & BLLac & 84.71 & -44.08 & 0.896 \\
\hline{4C +14.23} & {J0725.3+1426} & FSRQ & 111.33 & 14.44 & 1.038 \\
\hline{PMNJ 0531-4827} & {J0532.0-4826} & UNID & 83.01 & -48.44 &  $/$ \\
\hline{PKS 0402-362} & {J0403.9-3604} & BLLac & 60.99 & -36.07 & 1.417 \\
\hline{PKS 1124-186} & {J1126.6-1856} & FSRQ & 171.66 & -18.95 & 1.048 \\
\hline{Ton 599} & {J1159.5+2914} & FSRQ & 179.88 & 29.25 & 0.725 \\
\hline{PKS 2142-75} & {J2147.4-7534} & FSRQ & 326.87 & -75.58 & 1.139 \\
\hline{PKS 0208-512} & {J0210.8-5100} & FSRQ & 32.70 & -51.2 & 1.003 \\
\hline{PKS 0235-618} & {J0237.1-6136} & FSRQ & 39.29 & -61.62 & 0.467 \\
\hline{PKS 1830-211} & {J1833.6-2104} & FSRQ & 278.41 & -21.08 & 2,507 \\
\hline{PKS 2023-07} & {J2025.6-0736} & FSRQ & 306.42 & -7.61 & 1.388 \\
\hline{PKSB 1424-418} & {LJ1428.0-4206} & FSRQ & 217.01 & -42.10 & 1.522 \\
\hline{PMNJ 2345-1555} & {J2345.0-1553} & FSRQ & 356.27 & -15.89 & 0.621 \\
\hline{OJ 287} & {J0855.4+2009} & BLLac & 133.85 & 20.09 & 0.306 \\
\hline{PKS 0440-00} & {J0442.7-0017} & FSRQ & 70.69 & -0.29 & 0.845 \\
\hline{PKS 0250-225} & {J0252.7-2218} & FSRQ & 43.20 & -22.31 & 1.419 \\
\hline{B22308+34} & {J2311.0+3425} & FSRQ & 347.77 & 34.43 & 1.187 \\
\hline{B21520+31} & {J1522.1+3144} & FSRQ & 230.54 & 31.74 & 1.487 \\
\hline{PKS 1730-13} & {J1733.1-1307} & FSRQ & 263.28 & -13.13 & 0.902 \\
\hline{PKS0244-470} & {J0245.9-4652} & FSRQ & 41.06 & -47.06 & 1.385 \\
\hline{PKS 0301-243} & {J0303.4-2407} & BLLac & 45.87 & -24.13 & 0.26 \\
\hline{CTA 102} & {J2232.4+1143} & FSRQ & 338.12 & 11.72 & 1.037 \\
\hline{OG 050} & {J0532.7+0733} & FSRQ & 83.19 & 7.56 & 1.254 \\
\hline{PMNJ 2331-2148} & {J2330.9-2144} & FSRQ & 352.75 & -21.74 & 0.563 \\
\hline{PKS0805-07} & {J0808-0751} & BLLac & 122.06 & -7.85 & 1.837 \\
\hline{PKS 2320-035} & {J2323.6-0316} & FSRQ & 350.91 & -3.28 & 1.411 \\
\hline{PKS 2227-08} & {J2229.7-0832} & FSRQ  & 337.44 & -8.55 & 1.560 \\
\hline{OJ248} & {J0830.5+2407} & FSRQ  & 127.72 & 24.18 & 0.942 \\
\hline{PKS 2233-148} & {J2236.5-1431} & BLLac & 339.13 & -14.53 & 0.325 \\
\hline
\end{tabular}
\end{table}

\begin{figure}[ht!]
\centering
\includegraphics[width=0.9\textwidth]{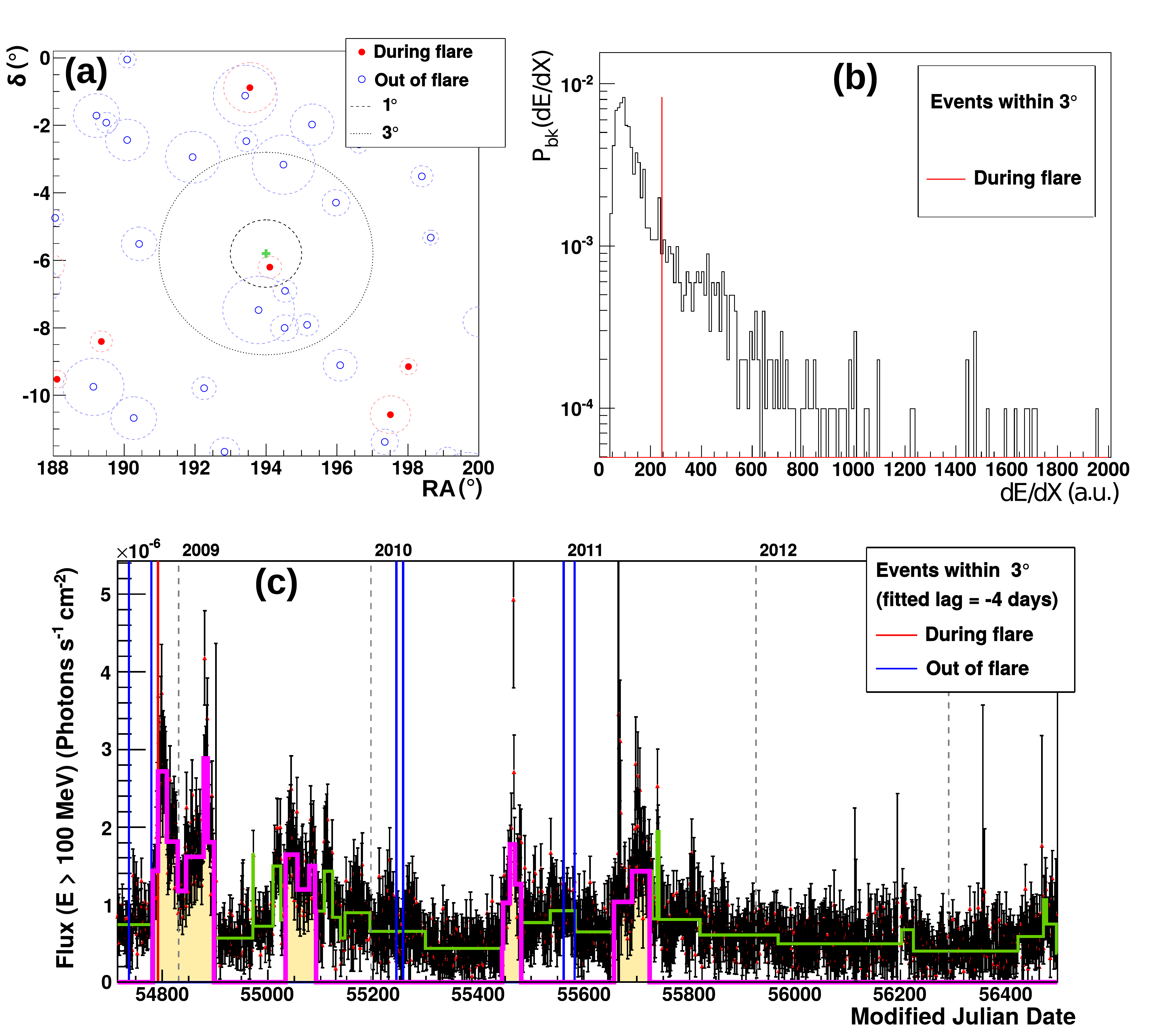}
\caption{Results for the blazar 3C279. (a) Event map around the direction of 3C279 indicated by the green cross. The full red
(hollow blue) dots indicate the events (not) in time coincidence with the selected flares. The size of the circle around the dots is proportional 
to the estimated angular uncertainty for each event. (b) Distribution of the energy estimator dE/dX in a $\pm10^{\circ}$ declination band around the 
source direction. The red line displays the value of the event in coincidence with the flare in a 1$^\circ$ cone around the source direction. (c) Gamma-ray light 
curve (red dots) with the error bars (black) measured by the LAT instrument above 100 MeV. The green and purple histograms show the denoised light curve and the 
selected flare periods respectively. The red and blue lines display the times of the ANTARES events associated with the source during a flaring state and other 
events in a 3$^\circ$ box around the source position, respectively.
}
\label{fig:3C279results1}
\end{figure}

The results of the search is summarised in Table~\ref{table:results}. Only three sources, 3C279, PKS10235-618 and PKS1124-186, have a pre-trial p-value lower than 10\%. The 
lowest p-value, 3.3\%, is obtained for 3C279 where one event is coincident with a 
large gamma-ray flare detected by Fermi/LAT in November 2008. Figure~\ref{fig:3C279results1} shows 
the Fermi gamma-ray light curve of 3C279 with the time of the neutrino events, the estimated energy
distribution, and the angular distribution of the events around the position of this source. This coincident event is reconstructed with 89 hits 
(energy deposition $dE/dX=244$ in arbitrary units) distributed on ten lines 
with a track fit quality $\Lambda=-4.4$. The particle track direction is reconstructed at 0.3$^{\circ}$ from 
the source location. This event has already been reported in the previous analysis~\cite{bib:flare}. The post-trial probability, computed by taking 
into account the 41 searches, is 67$\%$, and is thus compatible with background fluctuations.

In the absence of a discovery, upper limits on the neutrino fluence, $\mathcal{F}_{\nu}$, at 90\% confidence level are computed using 
5-95\% of the energy range as:

\begin{equation*}
  \mathcal{F_\nu} = \Delta t \int_{E_{\rm min}}^{E_{\rm max}} \mathrm{d}E \cdot E \frac{\mathrm{d}N}{\mathrm{d}E}
\end{equation*}

\begin{figure}[ht!]
\centering
\includegraphics[width=0.9\textwidth]{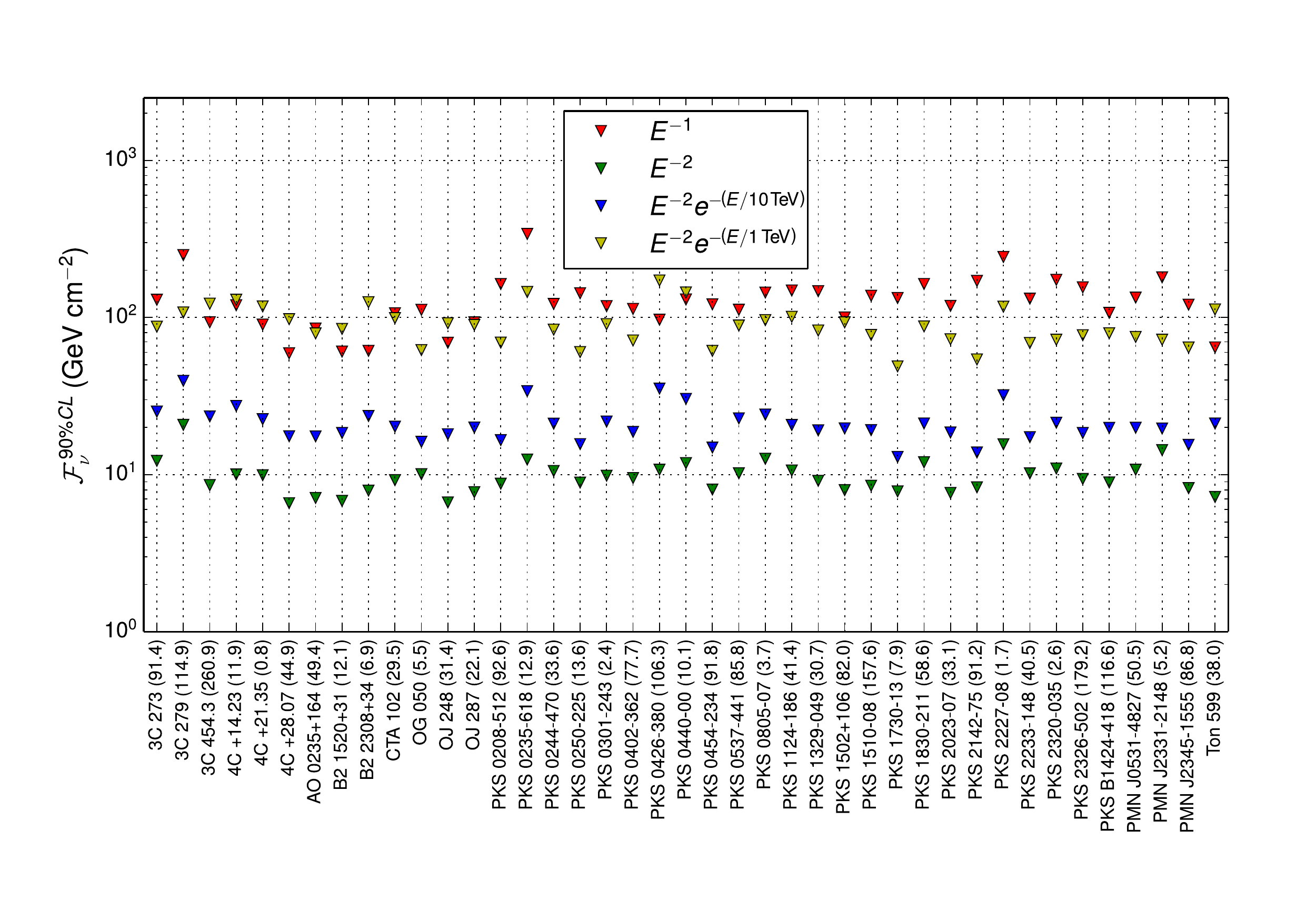}
\caption{Upper limits on the neutrino fluence for the 41 studied Fermi blazars in the case of  $E^{-2}$ (green), $E^{-2}\exp(-E/10~\rm{TeV})$ (blue), 
$E^{-2}\exp(-E/1~\rm{TeV})$ (yellow) and $E^{-1}$ (red) neutrino energy spectra. The number in parenthesis after the name of the source in the x-axis indicates 
the total effective flare duration $\Delta t$ during the studied period.
}
\label{fig:limit}
\end{figure}

The emission duration, $\Delta t$, is computed using the time PDF and the effective livetime. The limits include the systematic errors and are calculated according to 
the classical (frequentist) method for upper limits~\cite{bib:Neyman}. Figure~\ref{fig:limit} gives these upper limits. IceCube has performed a similar time-dependent 
analysis~\cite{bib:IceCubeflare} using data from 2008 to 2012, with 19 sources in common with the list presented in Table~\ref{table:Sources}. For 
sources in the Southern Hemisphere, the limits computed by IceCube are of the same order of magnitude as the ANTARES limits while they are a factor 10 stronger 
for the sources in the Northen hemisphere thanks to IceCube's larger detector volume.

\begin{table}[ht!]
\caption{Results of the search for neutrinos in coincidence with Fermi blazar flares (three first rows) and TeV flares (last row). The total duration of all
identified flares $\Delta t$, the optimised $\Lambda_{opt}$ cuts, the number of 
required events for a $3\sigma$ 
discovery ($N_{3\sigma}$ ), the number of fitted signal events by the likelihood ($N_{fit}$), the fitted time lag (Lag) and the corresponding pre-trial (post-trial) 
probability are given together with the energy spectra.}	       
\label{table:results}	 
\centering
\begin{tabular}{|c|c|c|c|c|c|c|c|c|}
\hline{Source}              &  	$\Delta t$	 &   $\Lambda_{opt}$ 	&    $N_{3\sigma}$ & $N_{fit}$ & Lag & P-value & Post-trial & Spectrum \\	      
\hline
\hline{3C279}               &  279 d  		 &    -5.3          	&    2.5   		&    0.8 & -4 d &  0.033 & 0.67 &      $E^{-2}$              	\\
\hline{PKS1124-186}         &  73 d 		 &    -5.4          	&    3.1   		&    0.7 & +4 d &  0.059 & 0.94 &      $E^{-2}\exp(\rm -E/1TeV)$             	\\
\hline{PKS0235-618}         &  25 d  		 &    -5.7          	&    1.5   		&    0.6 & -4 d &  0.045 & 0.91 &      $E^{-2}\exp(\rm -E/10TeV)$              	\\			
\hline
\hline{PKS0447-439}   &  10 d		 &    -5.4		&    0.75		&    0.1 & +5 d &  0.10 & 0.55 &      $E^{-2}\exp(\rm -E/1TeV)$			  \\
\hline
\end{tabular}
\end{table}

\section{Search for neutrino emission from gamma-ray flares detected by TeV telescopes}

Ground-based observatories such as H.E.S.S., MAGIC and VERITAS cannot monitor sources continuously, 
because they generally have a reduced field of view and a low duty cycle (e.g. only moonless nights). Nevertheless, these telescopes 
detect photons with energies from a few hundred GeV to a few TeV that may be better correlated 
with the neutrinos to which ANTARES is sensitive. These observatories often emit alerts reporting 
flares to Astronomer's Telegram or directly in a dedicated paper. When the start and stop times of the 
flare are known, the time PDF is assumed to be a single square-shaped flare with a minimum width of one 
day. Often, the beginning and the end of the flaring activity cannot be constrained accurately. In this 
case, a simple time cut is used, taking a time window including two days before and after 
the identified flare. Table~\ref{table:SourcesTeV} presents the list of seven TeV flares, their characteristics, 
and the publications from where this information is extracted. The flares are chosen for this analysis according to the same 
visibility criteria as for Fermi/LAT observations. The same analysis as described previously is performed 
assuming the same four energy spectra.

\begin{figure}[ht!]
\centering
\includegraphics[width=0.8\textwidth]{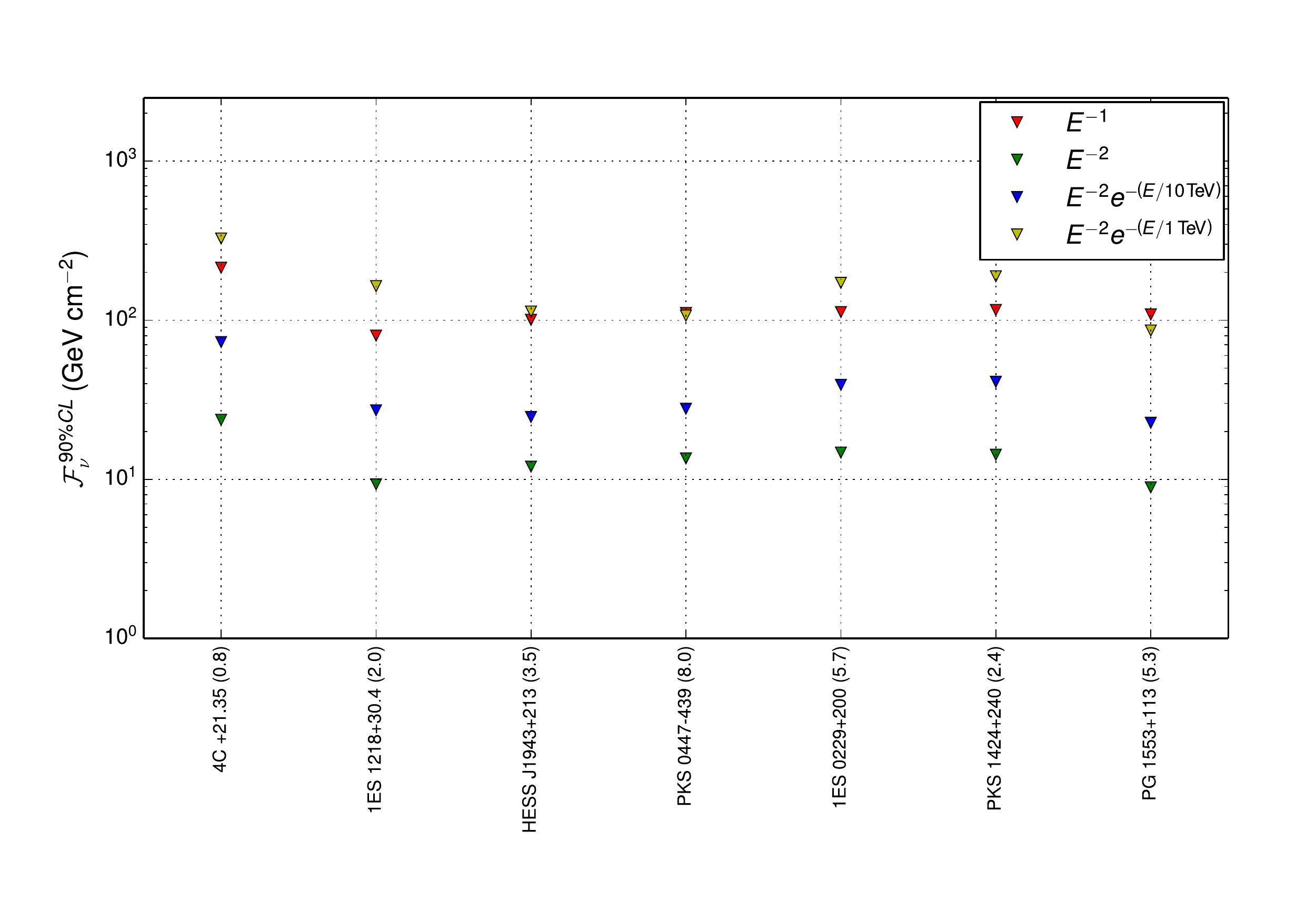}
\caption{Upper-limits on the neutrino fluence for the seven studied TeV blazars in the case of  $E^{-2}$ (green), $E^{-2}\exp(-E/10~\rm{TeV})$ (blue), 
$E^{-2}\exp(-E/1~\rm{TeV})$ (yellow) and $E^{-1}$ (red) neutrino energy spectra. The number in parenthesis after the name of the source in the x-axis indicates 
the total effective flare duration $\Delta t$ during the studied period.
}
\label{fig:limitsTeV}
\end{figure}

Six of the seven flares tested show no excess of events in the vicinity of the corresponding sources 
in the selected time windows. Only the blazar PKS0447-439 shows a pre-trial p-value lower than 10\% in the case 
of the assumed $E^{-2}\exp(-E/1~\rm{TeV})$ energy spectrum (cf Table~\ref{table:results}). The corresponding post trial p-value is 
55\%, and is also consistent with background fluctuations. In the absence of a signal, upper limits on the neutrino fluence at 90\% confidence 
level are computed  including the systematic errors (Figure~\ref{fig:limitsTeV}).

\begin{table}[ht!]
\caption{List of TeV flares reported by the three telescopes H.E.S.S., MAGIC and VERITAS in the 2008-2012 period.}	       
\label{table:SourcesTeV}	 
\centering
\begin{tabular}{|c ||c |c |c |c |c |}
\hline Name & Telescope & RA & Dec  &  Flaring days (MJD) & Reference \\
\hline
\hline{4C+21.35} & MAGIC & 186.2 & 21.4 & 55364-5 & arXiv:1101.4645  \\
\hline{PG 1553+113} & MAGIC & 239.0 & 11.2 & 55980-91 56037-8 & arXiv:1109.5860  \\
\hline{PKS 1424+240} & MAGIC & 216.8 & 23.8 & 54940-60 & arXiv:1109.5860  \\
\hline{1ES 1218+30.4} & VERITAS & 185.4 & 30.2 & 54860-5 & arXiv:1005.3747  \\
\hline{1ES 0229+200} & VERITAS & 38.2 & 20.3 & 55118-31 & arXiv:1307.8091  \\
\hline{J1943+213} & H.E.S.S. & 296.0 & 21.3 & 55040-60 & arXiv:1103.0763  \\
\hline{PKS0447-439} & H.E.S.S. & 72.4 & -43.8 & 55174-84 & arXiv:1303.1628  \\
\hline
\end{tabular}
\end{table}

\section{Discussion}
\begin{figure}[ht!]
\centering
\includegraphics[width=0.9\textwidth]{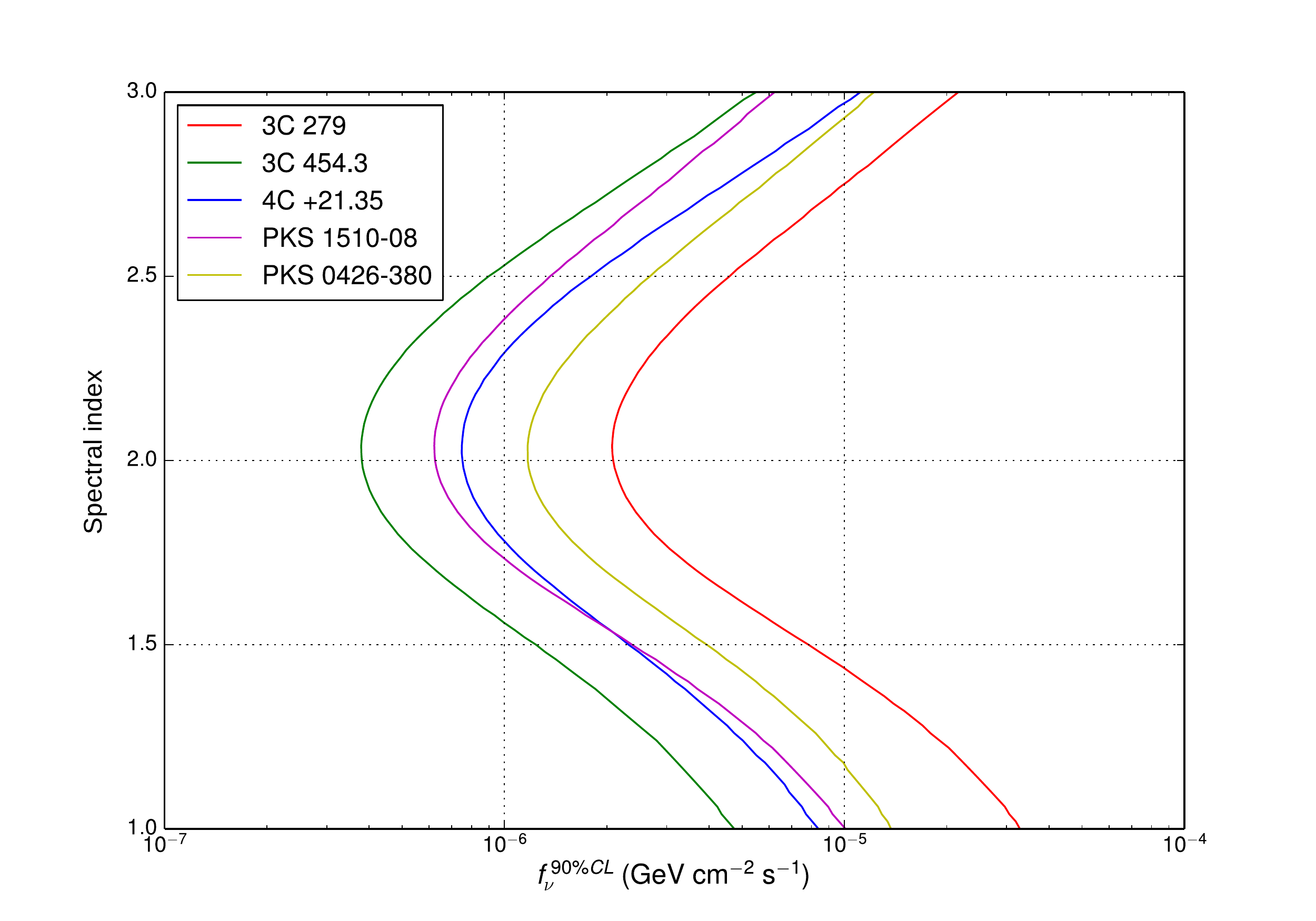}
\caption{Extrapolation of the neutrino upper limit energy fluxes (integrated in the 5-95\% energy range) from E$^{-2}$ to other spectral indices for the five brightest Fermi blazars.
}
\label{fig:limitvsindex}
\end{figure}

%Hadronic interactions $p-\gamma$ or p-p
Hadronic interactions predict neutrino emission in the TeV-PeV range associated with a flux of gamma rays. The prediction that the total neutrino energy flux $F_{\nu}$ 
is approximately equal to the total high-energy photon flux $F_{\gamma}$ is relatively robust, at least when attributing this emission to a 100\% hadronic origin~\cite{bib:Kelner,bib:KA}. 
As it is not affected by pair production losses, the neutrino emission is expected to have energies systematically higher than the electromagnetic component. The 
Fermi measurement of the source 
flux in the 0.1-100~GeV range is an underestimate of the overall electromagnetic spectrum, since it covers only three decades of energy. This underestimates, 
around a factor of 2-3~\cite{bib:tchernin}, has to be taken into account in the comparison between neutrinos and gamma-rays. Moreover, the lack of TeV 
observations makes a direct comparison at the >TeV range difficult, and necessitates an extrapolation 
of the LAT data over several orders of 
magnitude. This extrapolation is performed using a fit of the Fermi data with a log-parabola or a power-law function, both of which have been used 
by the Fermi Collaboration to build its source catalogues. To 
see how the neutrino upper limits and photon energetics compare, the gamma-ray spectral energy distributions (SEDs) of all sources have been produced 
using the SED builder of the ASI Science Data Centre (ASDC)~\cite{bib:Asdc} adding, if needed, VHE data taken from the literature, building a hybrid photon-neutrino SED~\cite{bib:Resconi}. The limits for different spectral 
indices, from -3 to -1, 
are extrapolated from the limits obtained in the $E^{-2}$ energy spectra using the corresponding acceptance curve. Figure~\ref{fig:limitvsindex} shows the 
value of the neutrino flux limits for the five brighest sources as a function of the spectral indices. The range in energy for each limit corresponds 
to the 5-95\% range of the energy distribution for a given spectra.

\begin{figure}[ht!]
\centering
\includegraphics[width=0.9\textwidth]{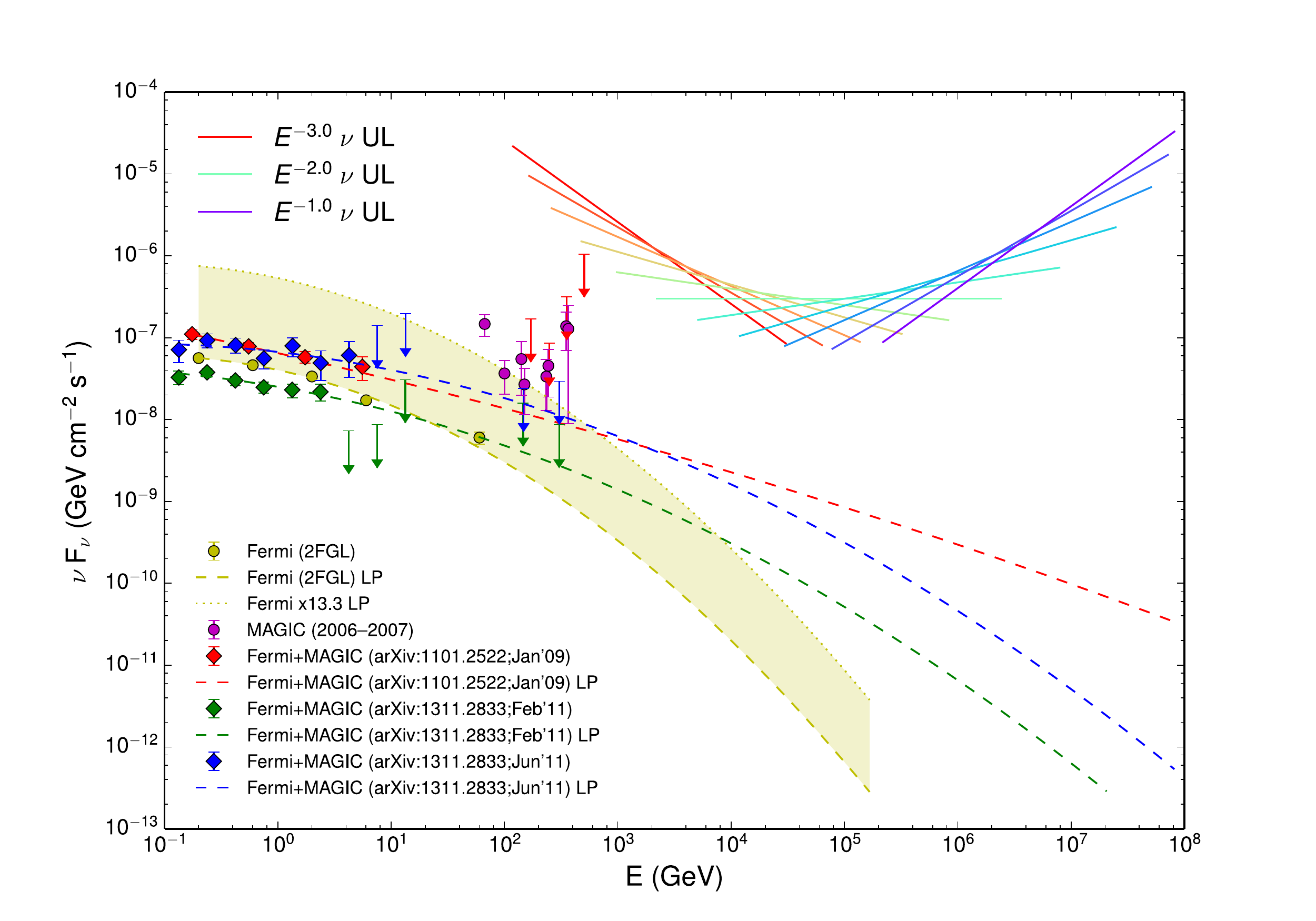}
\caption{Gamma-ray SED of 3C279, observed by Fermi/LAT and MAGIC in January 2009 (red dots), January 2011 (green dots) and June 2011 (blue dots)~\cite{bib:3c279Magic1,bib:3c279Magic2}. 
The yellow/green dots are the average flux with 2008-2010 data (2FGL). The observed data points are corrected 
for absorption by the extragalactic background light assuming z = 0.536. The dashed lines represents the fits performed on Fermi data using a log-parabola (LP) function. 
The shaded yellow area represents an extrapolation of the flux during the
studied flares from the average flux observed by Fermi. The lower bound is computed from the average flux during 2008-2012 period, while the upper bound is the maximum 
flux measured in the light curve (x13.3 the average flux). 
Finally, the coloured solid lines indicates the neutrino upper limits for different spectral indices (from $E^{-3}$ in red to $E^{-1}$ in blue).
}
\label{fig:hybridSED_3C279}
\end{figure}

Figure~\ref{fig:hybridSED_3C279} shows the hybrid SED for the lowest p-value source 3C279. The shaded yellow area represents an extrapolation of the flux during the
studied flares from the average flux observed by Fermi. The lower bound is computed from the average flux during the 2008-2012 period, while the upper bound is simply the renormalised flux according to
the maximum flux measured in the light curve. With this simple criteria of the energy budget, the limit set by ANTARES for 
the blazar 3C279 is of the same order of magnitude as the electromagnetic flux measurements during the flares. It reinforces the need to search for a neutrino 
signal during the outburst periods when the gamma-ray flux and the accompanying neutrino flux are much higher. Therefore, with more data, ANTARES should be able 
to significantly constrain a 100$\%$ hadronic origin of the high-energy gamma-ray emission. Fermi has reported some very intense outburst periods of 3C279 between 
mid 2013 and end of 2014~\cite{bib:3c279atel1,bib:3c279atel2}, periods not considered in this paper.

\begin{figure}[ht!]
\centering
\includegraphics[width=0.9\textwidth]{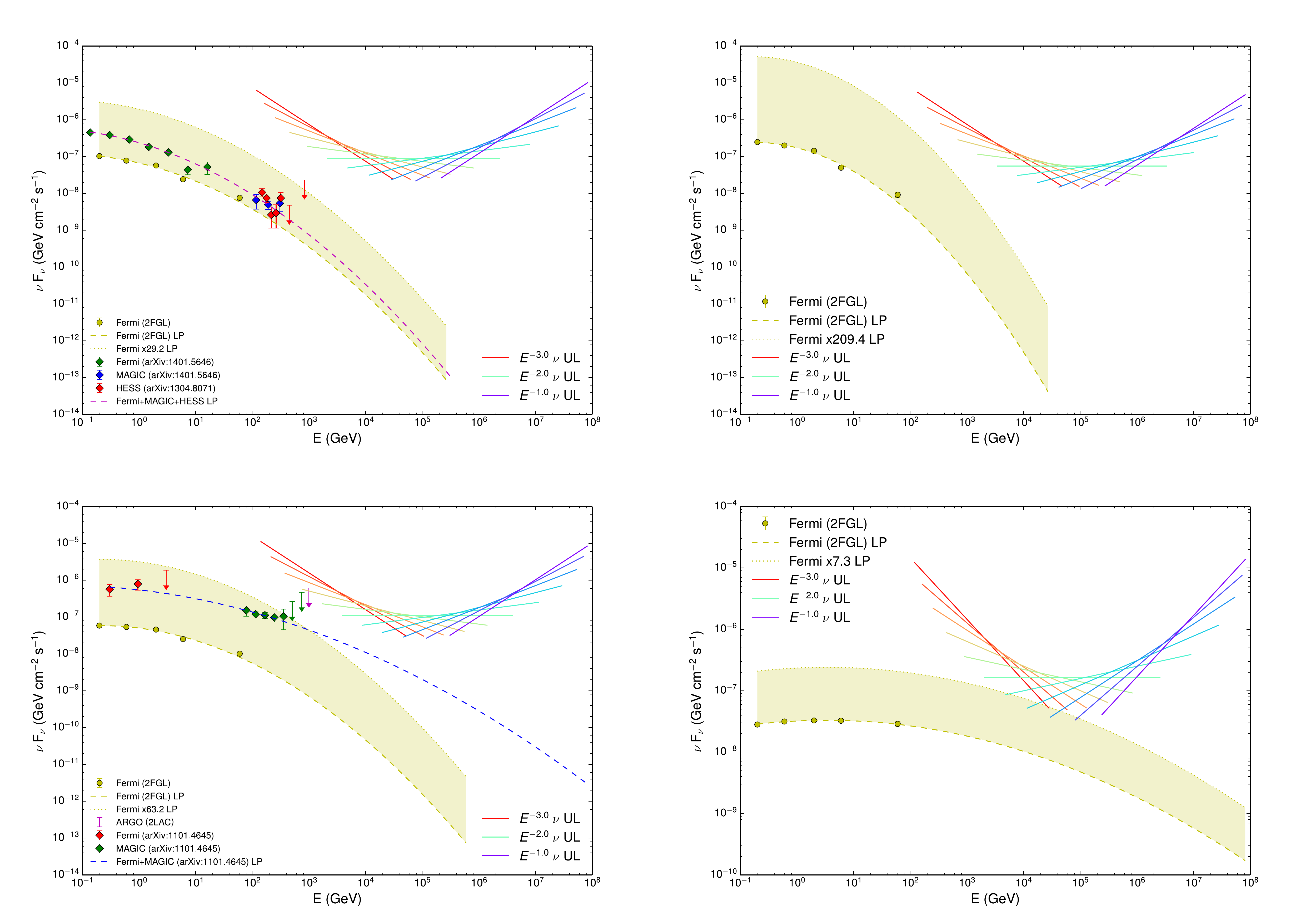}
\caption{Hybrid SED of the four additional bright Fermi blazars  PKS1510-089, 3C454.3, 4C21+35 and PKS0426. Points and lines are as described in the caption to Figure~\ref{fig:hybridSED_3C279}.
}
\label{fig:hybridSED_others}
\end{figure}

Figure~\ref{fig:hybridSED_others} shows the hybrid SEDs for the four additional bright blazars. These sources, classified as FSRQs, have their 
gamma-ray flux suppressed at high energy. Only the brightest FSRQ objects are detected at TeV energies. On 
the contrary, BL Lac objects have high-energy spectral components, favoured for TeV gamma-ray detection, but are fainter than FSRQs. Both 
FSRQs~\cite{bib:FSRQneut,bib:FSRQneut2} and BL Lac objects~\cite{bib:BLLacsneut} have been argued to be potential sources of neutrinos. This 
comparison between the gamma-ray flux and the neutrino flux limit provides an indication as to how to build an optimised source list for future 
searches with ANTARES and its successor KM3NeT~\cite{bib:KM3NET}.

\section{Summary}
This paper discusses the extended time-dependent search for cosmic neutrinos using the data taken with the full ANTARES 
detector between 2008 and 2012. For variable sources, time-dependent searches are significantly more sensitive than steady 
point-source searches thanks to the large reduction of the atmospheric background over short time scales. 
These searches have been applied to 41 very bright and variable Fermi LAT blazars and 
seven TeV flares reported by H.E.S.S, VERITAS or MAGIC telescopes. The most significant correlation was found with a GeV flare of the 
blazar 3C279 for which one neutrino event was detected in time/spatial coincidence with the gamma-ray emission. However,
this event has a post-trial probability of 67$\%$, and is thus compatible with background fluctuations. Upper limits were obtained 
on the neutrino fluence for the selected sources and compared with the high-energy component of the spectral energy distributions 
computed with GeV-TeV gamma-ray observations. These comparisons show that for the brighter blazars, the neutrino flux limits are of 
the same order of magnitude as the high-energy gamma-ray fluxes. With additional data from ANTARES and with the order of magnitude 
sensitivity improvement expected from the next generation neutrino 
telescope, KM3NeT, the prospects for future searches for neutrino emission from flaring blazars are very promising.

\acknowledgments

The authors acknowledge the financial support of the funding agencies:
Centre National de la Recherche Scientifique (CNRS), Commissariat \`a
l'\'ener\-gie atomique et aux \'energies alternatives (CEA),
Commission Europ\'eenne (FEDER fund and Marie Curie Program), R\'egion
\^Ile-de-France (DIM-ACAV) R\'egion Alsace (contrat CPER), R\'egion
Provence-Alpes-C\^ote d'Azur, D\'e\-par\-tement du Var and Ville de La
Seyne-sur-Mer, France; Bundesministerium f\"ur Bildung und Forschung
(BMBF), Germany; Istituto Nazionale di Fisica Nucleare (INFN), Italy;
Stichting voor Fundamenteel Onderzoek der Materie (FOM), Nederlandse
organisatie voor Wetenschappelijk Onderzoek (NWO), the Netherlands;
Council of the President of the Russian Federation for young
scientists and leading scientific schools supporting grants, Russia;
National Authority for Scientific Research (ANCS), Romania; 
Mi\-nis\-te\-rio de Econom\'{\i}a y Competitividad (MINECO), Prometeo 
and Grisol\'{\i}a programs of Generalitat Valenciana and MultiDark, 
Spain; Agence de  l'Oriental and CNRST, Morocco. We also acknowledge 
the technical support of Ifremer, AIM and Foselev Marine for the sea 
operation and the CC-IN2P3 for the computing facilities.

% The bibliography will probably be heavily edited during typesetting.
% We'll parse it and, using the arxiv number or the journal data, will
% query inspire, trying to verify the data (this will probalby spot
% eventual typos) and retrive the document DOI and eventual errata.
% We however suggest to always provide author, title and journal data:
% in short all the informations that clearly identify a document.

\end{document}